\documentclass[notitlepage]{revtex4-1} 
\usepackage{graphicx}

\usepackage{hyperref} 					 
\usepackage{listings} 					 
\usepackage{color}    					 
\usepackage{natbib}   					 
\usepackage{subfig} 					 

\usepackage{booktabs}
\usepackage[ansinew]{inputenc}				     
\usepackage{psfrag}
\usepackage{pstricks}
\usepackage{amsfonts}
\usepackage{amsmath}
\usepackage{bm}
\usepackage{lipsum}
\usepackage{pgf}
\usepackage{tikz}
\usetikzlibrary{patterns}

\usepackage{array}
\newcolumntype{C}[1]{>{\centering\arraybackslash}m{#1}}

\usepackage{algpseudocode}
\usepackage{color}

\usepackage{filecontents}

\begin{document}

\title{Inverse Design of Nanoparticles for Enhanced Raman Scattering}

\author{Rasmus E. Christiansen,$^{1,2}$}  
\email{Corresponding email: raelch@mek.dtu.dk}

\author{ J\'{e}r\^{o}me Michon,$^{3}$} \author{Mohammed Benzaouia,$^{4}$} \author{Ole Sigmund,$^{1}$} \author{Steven G. Johnson.$^{2}$}

\affiliation{$^{1}$ Department of Mechanical Engineering, Technical University of Denmark, Nils Koppels All\'{e}, Building 404, 2800 Kongens Lyngby, Denmark}
\affiliation{$^{2}$ Department of Mathematics, Massachusetts Institute of Technology, Cambridge, MA 02139, USA}
\affiliation{$^{3}$Department of Materials Science and Engineering, Massachusetts Institute of Technology, Cambridge, MA 02139, USA}
\affiliation{$^{4}$ Department of Electrical Engineering and Computer Science, Massachusetts Institute of Technology, Cambridge, MA 02139, USA}
%



\begin{abstract}
We show that topology optimization (TO) of metallic resonators can lead to $\sim 10^2\times$ improvement in surface-enhanced Raman scattering (SERS) efficiency compared to traditional resonant structures such as bowtie antennas. TO inverse design leads to surprising structures very different from conventional designs, which simultaneously optimize focusing of the incident wave and emission from the Raman dipole. We consider isolated metallic particles as well as more complicated configurations such as periodic surfaces or resonators coupled to dielectric waveguides, and the benefits of TO are even greater in the latter case. Our results are motivated by recent rigorous upper bounds to Raman scattering enhancement, and shed light on the extent to which these bounds are achievable.
\end{abstract}

\maketitle


\section{Introduction}
\begin{figure}[h!]
	\centering\includegraphics[width=0.7\linewidth]{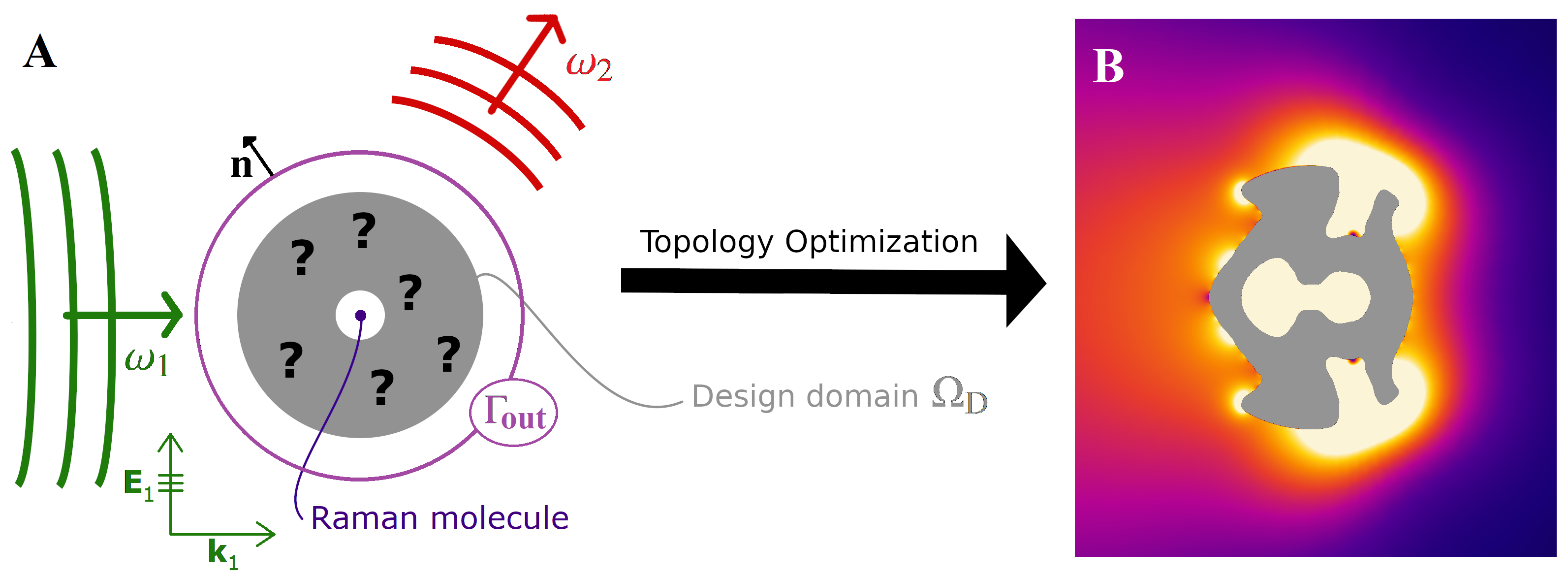}
	\caption{ \textbf{A)} Sketch of the Raman scattering design problem. \textbf{B)} $\vert \textbf{E}_2 \vert$-field emitted through Raman scattering from a molecule (point dipole) placed at the centre of a (gray) topology-optimized silver structure at $\lambda_1 = \lambda_2 = 532$ nm. The color scheme is saturated due to the diverging field at the source.  \label{FIG:RAMAN_PROBLEM_ILLUSTRATION}}
\end{figure}

In this paper, we present freeform shape optimization ("topology optimization," TO \cite{BOOK_TOPOPT_BENDSOE,JENSEN_SIGMUND_2011,MOLESKY_2018}) of metallic nanoparticles for Raman scattering \cite{Long1977,Turrell1996,Colthup1990}, and obtain non-intuitive structures $\sim 60\times$ better (in terms of emitted power) than optimized coupled-sphere \cite{Huang2007,Zhu2011,Rechberger2003} or bowtie \cite{Hao2004,Dodson2013,Kaniber2016,Yue2017,Zhang2015} antennas (Sec.~\ref{SEC:REFERENCE_DESIGNS}) for identical separation distance to the Raman molecule. Our current results are a proof-of-concept of Raman TO in 2d systems, and the resulting dramatic enhancements suggest exciting opportunities for future improvements in practical 3d Raman sensing. Stated briefly, Raman scattering consists of an incident wave at a frequency $\omega_1$ interacting with a molecule that subsequently emits electromagnetic radiation at $\omega_2$. A nanostructure can enhance both the incident-wave absorption by a focusing effect as well as the emission by a Purcell effect \cite{Moskovits1985,Campion1998,Kneipp2006,leru2008,Stiles2008,Haynes2015}. Figure~\ref{FIG:RAMAN_PROBLEM_ILLUSTRATION}A shows a schematic of this process, in which the molecule is surrounded by an unknown arrangement of metal (e.g. silver); TO is used to tailor this arrangement to maximize the Raman scattering (Sec.~2), resulting in surprising structures such as the one shown in Fig.~\ref{FIG:RAMAN_PROBLEM_ILLUSTRATION}B. In addition to optimizing isolated resonators coupled with incident planewaves (Fig.~\ref{FIG:RAMAN_PROBLEM_ILLUSTRATION}A), we also optimize Raman scattering on periodic surfaces as well as resonators coupled with input/output waveguides, as depicted in  Fig.~\ref{FIG:MODEL_PROBLEMS}B and Fig.~\ref{FIG:MODEL_PROBLEMS}C, respectively. We show that it is important to consider the full Raman process combining \emph{both} focusing and emission, as optimizing either emission or focusing alone sacrifices a factor of $\sim 5\times$ (Sec.~\ref{SEC:DIFFERENT_OBJECTIVES}). When the Raman shift $\omega_1 - \omega_2$ is more than a few percent (i.e., comparable to the bandwidth of a single plasmonic resonance), we show that this shift must be taken into account in the optimization, or one may sacrifice a factor of $\sim 5\times$ (Sec.~\ref{SEC:RAMAN_SHIFT}). We find that the huge enhancements observed from TO structures compared to simple geometries are in qualitative agreement with recently discovered theoretical upper bounds to Raman scattering \cite{MICHON_ET_AL_2019}. Quantitatively, for a material with susceptibility $\chi$, the key figure of merit for light-matter interactions is $F=\vert \chi \vert ^2/\mathrm{Im}(\chi)$ \cite{Miller2016,Averitt1999,Bohren1998} and the Raman bounds scale as a factor of $F^3 V$ where $V$ is the volume of the scatterer: a factor of
$F^2$ maximum focusing enhancement and a factor of $F$ maximum emission enhancement. The enhancement achieved by our TO designs scale roughly linearly with $V$ in agreement with the upper bounds (Sec.~\ref{SEC:SIZE_SCALING}), but they scale roughly with $F^2$ instead of $F^3$ (Sec.~\ref{SEC:MATERIAL_SCALING}) suggesting that in practice one cannot simultaneously achieve ideal focusing and emission enhancement.

In conventional Raman spectroscopy, the very small Raman cross-section of most chemicals results in Raman radiation typically on the order of 0.001\% of the power of the pump signal \cite{Long1977}. This low efficiency is overcome in surface-enhanced Raman spectroscopy (SERS) by placing the chemicals of interest in the vicinity of a scatterer, typically a surface or collection of nanostructures, which increases the overall signal that can be collected \cite{Moskovits1985,Campion1998,Kneipp2006,leru2008,Stiles2008,Haynes2015}. This technique has allowed for efficiencies up to 12 orders of magnitude larger than that of traditional Raman spectroscopy, yielding detection levels down to the single molecule \cite{Nie1997,Kneipp1997} and opening up applications in the fields of biochemistry, forensics, food safety, threat detection, and medical diagnostics \cite{Stiles2008,Haynes2015}. While many different materials and antenna geometries have been used for SERS substrates \cite{Sharma2012,Fateixa2015,Mosier2017}, so far the optimization of these geometries has been limited to one or two degrees of freedom in designs such as bowtie antennas \cite{Camden2008,LeRu2007,Hao2004,Genov2004,Sundaramurthy2005}. Comparison with the recently derived upper bound to Raman enhancement showed these geometries to be greatly suboptimal, with performance several orders of magnitude below the bounds \cite{MICHON_ET_AL_2019}. Although it is possible that the bounds can be tightened (e.g. by incorporating additional physical constraints \cite{MOLESKY_2019}), the fact that current SERS geometries are so far below these new theoretical limits made us wonder if dramatic improvements might be attainable by TO.

In this work we thus take a hitherto unexplored approach in the context of Raman scattering, in which the problem of designing metallic nanostructures to enhance the process is formulated as a mathematical optimization problem and solved using density-based topology optimization \cite{BOOK_TOPOPT_BENDSOE}. In the design process, we simultaneously optimize the focusing of the incident field and the emission from the molecule (dipole), which is shown to lead to structures with higher performance than only optimizing for one process. In brief, density-based topology optimization operates by introducing a continuous design field to control the physical material distribution, enabling the use of adjoint sensitivity analysis \cite{TORTORELLI_ET_AL_1994} and gradient-based optimization algorithms \cite{BOOK_OPTIMIZATION_NOCEDAL,SVANBERG_2002} to efficiently solve design problems with potentially billions of design degrees of freedom \cite{AAGE_ET_AL_2017} . Hence, the approach provides near-unlimited design freedom, with a computational complexity dominated by the solution of the Maxwell equations, utilizing mature finite-element techniques \cite{BOOK_FEM_EM_JIN}. A suite of well-understood tools, developed or matured over the last decades, are used to solve the optimization problem, interpolate material parameters, control design variations and ensure physical admissibility of the design \cite{SVANBERG_2002,BOURDIN_ET_AL_2001,GUEST_ET_AL_2004,WANG_ET_AL_2011,CHRISTIANSEN_VESTER-PETERSEN_2019}. In addition, geometric length-scale constraints are employed to ensure that all features of the final designs are above a specified size (which may be chosen to comport with fabrication constraints) \cite{ZHOU_ET_AL_2015}. Over the past 20 years, TO has been applied to an increasingly wide range of problems in electromagnetic design \cite{JENSEN_SIGMUND_2011,MOLESKY_2018}. Our work on Raman optimization couples multi-frequency focusing and emission problems. Maximizing the emission alone (for a dipole source) corresponds to maximizing the local density of states (LDOS) \cite{FREI_ET_AL_2008,LU_ET_AL_2011}, a formulation that has been employed for TO of optical cavities \cite{LIANG_AND_JOHNSON_2013,WANG_2018}.  The focusing problem alone is related to lens and antenna design among others, for which TO has also been applied (both to small metallic particles such as those considered here and to much larger structures, such as metalenses) \cite{WADBRO_ET_AL_2015,DENG_ET_AL_2015,VESTER-PETERSEN_2017,VESTER-PETERSEN_2018,CHRISTIANSEN_VESTER-PETERSEN_2019,CHUNG_2019}. Nonlinearly coupled electromagnetic resonances at multiple frequencies were optimized via TO for second harmonic generation \cite{LIN_2016}.

\section{Model Formulation} \label{SEC:MODEL_PROBLEM}

The Raman scattering phenomenon (sketched in Fig.~\ref{FIG:RAMAN_PROBLEM_ILLUSTRATION}A) is modelled as two sequentially-coupled, time-harmonic, classical electromagnetic problems \cite{BOOK_EM_GRIFFITHS,BOOK_FEM_EM_JIN} in a domain $\boldmath{\Omega}$. The first models an in-plane polarized planewave propagating through the domain, interacting with any material distributed inside. The second models the excitation of a Raman molecule in $\boldmath{\Omega}$ by the electric field resulting from the first problem. The Raman molecule is modelled as a dipole source, with its dipole moment given by a polarizability tensor multiplied by the electric field, excited by the incident planewave at the position of the dipole \cite{leru2008}. The problem may be stated formally as

\begin{eqnarray}
\nabla \times \left(\mu^{-1} \nabla \times \textbf{E}_1(\textbf{r}) \right) - \omega_1^2 \varepsilon(\textbf{r}) \textbf{E}_1(\textbf{r})  = \textbf{f}_1(\textbf{r}), \ \ \ \textbf{r} \in \boldmath{\Omega},  \label{EQN:MAXWELL_PROBLEM1} \\
\textbf{f}_2(\textbf{r}) = - \omega_2^2 {\alpha} \textbf{E}_1(\textbf{r}_0) \delta(\textbf{r}-\textbf{r}_0),  \label{EQN:DIPOLE_PROBLEM2}  \\
\nabla \times \left(\mu^{-1} \nabla \times \textbf{E}_2(\textbf{r}) \right) - \omega_2^2 \varepsilon(\textbf{r}) \textbf{E}_2(\textbf{r}) = \textbf{f}_2(\textbf{r}), \ \ \ \textbf{r} \in \boldmath{\Omega}. \label{EQN:MAXWELL_PROBLEM2}
\end{eqnarray}

\noindent Here $\boldmath{\Omega} \in \mathbb{R}^2$ represents the modelling domain and $\textbf{r}$ and $\textbf{r}_0$ denote points in  $\boldmath{\Omega}$. The material parameters $\mu$ and $\varepsilon(\textbf{r})$ are the magnetic permeability and the electric permittivity, respectively (non-magnetic materials are assumed in the following, i.e. $\mu_r \equiv 1$). Further, $\omega_i = c / \lambda_i$ denotes the angular frequency with $c$ being the speed of light and $\lambda_i$ the wavelength, $\textbf{E}_i$ the electric field and $\textbf{f}_i$ the excitation sources for $i \in {1,2}$. Finally $\boldmath{\alpha}$ denotes the polarizability tensor, which in this work is taken to be the identity $\big(\alpha \textbf{E}_1(\textbf{r}_0) = \textbf{I} \textbf{E}_1(\textbf{r}_0)\big)$. 

The problem of enhancing the power emitted from the Raman molecule at $\lambda_2$, by tailoring the material distribution in the design domain $\boldmath{\Omega}_{\mathrm{D}}$, is formulated as a continuous optimization problem and solved using density-based topology optimization. In its basic form the optimization problem may be written as

\begin{eqnarray} \label{EQN:DESIGN}
&\underset{\xi(\textbf{r}) \in [0,1]}{\max} \ \ \ &\Phi(\xi) = \int_{\Gamma_{\mathrm{out}}} \left\langle  \textbf{S}_2(\xi) \right\rangle \cdot \textbf{n} \ \mathrm{d}\textbf{r}, \label{EQN:OBJECTIVE_FUNCTION} \\ \label{EQN:DESIGN_CON_1}
&\mathrm{s.t.} \ \ \ &0 \leq \xi(\textbf{r}) \leq 1, \ \textbf{r} \in \boldmath{\Omega}_{\mathrm{D}}.
\end{eqnarray}

\noindent Here $ \left\langle \textbf{S}_2 \right\rangle = \mathrm{Re}\left(\frac{1}{2} \textbf{E}_2 \times \textbf{H}_2^{*} \right)$ denotes the time averaged Poynting vector computed from $\textbf{E}_2$, which in turn is obtained by solving Eqs.~(\ref{EQN:MAXWELL_PROBLEM1})--(\ref{EQN:MAXWELL_PROBLEM2}) for a given $\xi(\textbf{r})$ and $(\bullet)^*$ denotes the complex conjugate. The vector $\textbf{n}$ is the outward pointing normal vector for the integration curve $\Gamma_{{\mathrm{out}}}$ (Fig.~\ref{FIG:RAMAN_PROBLEM_ILLUSTRATION}A). The permittivity distribution $\varepsilon(\textbf{r})$ is determined by the value of $\xi(\textbf{r})$ through the interpolation \cite{CHRISTIANSEN_VESTER-PETERSEN_2019},

\begin{eqnarray}
\varepsilon(x) &=& \big(n(x)^2 - \kappa(x)^2\big) - \mathrm{i} \big(2 n(x) \kappa(x)\big), \nonumber \\ 
\hspace{15pt} n(x) &=& n_{\mathrm{M}_1} + x (n_{\mathrm{M}_2} - n_{\mathrm{M}_1}), \\ \nonumber \hspace{15pt} \kappa(x) &=& \kappa_{\mathrm{M}_1} + x  (\kappa_{\mathrm{M}_2} - \kappa_{\mathrm{M}_1}).
\end{eqnarray}
 
\noindent Here $\mathrm{M}_i$ denotes the materials considered (metal or air) and $n$ and $\kappa$ denote the refractive index and extinction coefficient of $\mathrm{M}_i$, respectively. Finally, i denotes the imaginary unit.

Density-based topology optimization is used to solve Eqs.~(\ref{EQN:DESIGN})--(\ref{EQN:DESIGN_CON_1}) where the physical admissibility of the final material distributions is assured using filtering, thresholding and continuation techniques developed in the mechanics community \cite{WANG_ET_AL_2011,GUEST_ET_AL_2004}. This allows the enforcement of a minimum length scale (6 nm) on all features in the material distribution using geometric length-scale constraints \cite{ZHOU_ET_AL_2015}. 

\begin{figure}[h!]
	\centering\includegraphics[width=0.7\linewidth]{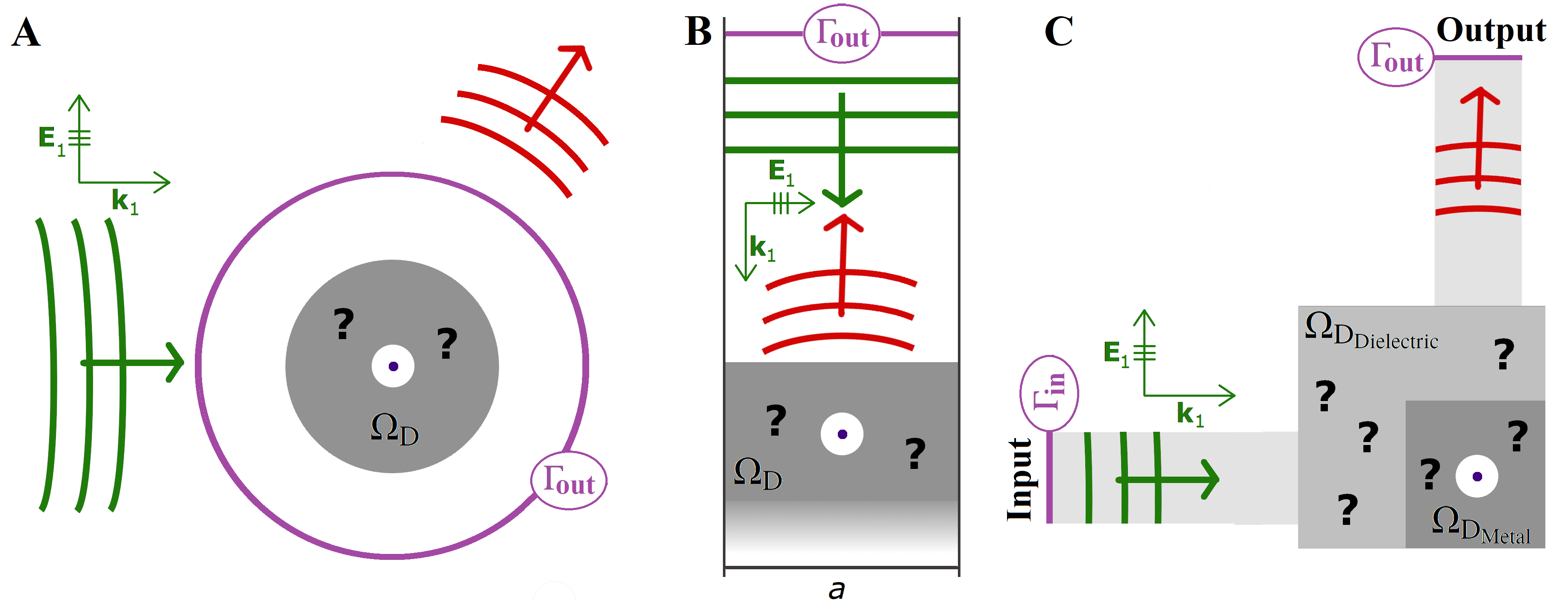}
	\caption{Problem configurations. \textbf{A)} A Raman molecule (blue) in air background, surrounded by the design domain $\boldmath{\Omega}_{\mathrm{D}}$ (gray), excited by an incident planewave (green), the total emitted power (red) through $\Gamma_{{\mathrm{out}}}$ is sought maximized. \textbf{B)} A Raman molecule positioned on a periodically patterned surface (period: $a$), surrounded by $\boldmath{\Omega}_{\mathrm{D}}$, excited by a planewave at normal incidence, maximizing the emitted power normal to the surface. \textbf{C)} A Raman molecule positioned inside $\boldmath{\Omega}_{\mathrm{D}}$, excited by an incident wave, coupled into the domain via a waveguide (input), maximizing the emitted power into another waveguide (output). \label{FIG:MODEL_PROBLEMS}}
\end{figure}

We consider the three configurations shown in Fig.~\ref{FIG:MODEL_PROBLEMS}. The first (Fig.~\ref{FIG:MODEL_PROBLEMS}A, Sec.~\ref{SEC:REFERENCE_DESIGNS}--\ref{SEC:WAVELENGTH_SCALING}) is used to benchmark the proposed approach. It consists of a Raman molecule (blue dot) placed in an air background with $\boldmath{\Omega}_{\mathrm{D}}$ (gray region) surrounding the molecule. An incident planewave propagates through the domain from left to right (green) and the power emitted through $\Gamma_{{\mathrm{out}}}$ by Raman scattering (red)  from the molecule is sought maximized. 

The second problem (Fig.~\ref{FIG:MODEL_PROBLEMS}B, Sec.~\ref{SEC:OUT_OF_PLANE_SCATTERING}) models out-of-plane Raman scattering and considers a metallic surface with periodic patterning placed in a dielectric medium. The model problem is $x$-periodic with period $a$. It consists of a spatial region containing the design domain with a Raman molecule placed at its centre, truncated from below by a metallic surface. An incident planewave propagates through the domain from top to bottom and the power emitted from the molecule is sought maximized in the direction normal to the metallic surface. The Raman scattering process from the different molecules on the surface is assumed to be incoherent. Therefore a $k$-space integration technique, the array scanning method \cite{SIGELMANN_ET_AL_1965,CAPOLINO_ET_AL_2007}, is used to compute the power emitted from each individual Raman molecule situated in the periodic background. 

The third model problem (Fig.~\ref{FIG:MODEL_PROBLEMS}C, Sec.~\ref{SEC:IN_PLANE_SCATTERING}) concerns in-plane Raman scattering into a waveguide (output). A Raman molecule is excited by an incident field coupled into the system from a second waveguide (input). The Raman molecule is positioned at the centre of $\boldmath{\Omega}_{\mathrm{D}_\mathrm{Metal}}$ (dark gray), in which a metallic nanostructure is designed. In the remaining part of the design domain, $\boldmath{\Omega}_{\mathrm{D}_\mathrm{Dielectric}}$ (light gray), a dielectric structure, aimed at coupling the light into and out of the waveguides, is designed simultaneously. For this problem, the difference in the power emitted by the Raman molecule through $\Gamma_{{\mathrm{out}}}$ and $\Gamma_{{\mathrm{in}}}$ is sought maximized.

The model domains are truncated using different boundary conditions depending on which problem configuration is considered. For the isolated resonators (Fig.~\ref{FIG:MODEL_PROBLEMS}A) and the resonators coupled with the input/output waveguides (Fig.~\ref{FIG:MODEL_PROBLEMS}C), a perfectly matched layer is used to truncate $\boldmath{\Omega}$. For the Raman molecules placed on a periodic surface (Fig.~\ref{FIG:MODEL_PROBLEMS}B), Floquet-Bloch boundary conditions are used in plane, a perfectly matched layer truncates the top boundary and a perfect electric conduction condition is imposed on the bottom boundary.

The model problems are implemented in COMSOL Multiphysics \cite{COMSOL54} and the optimization problem is solved using the Globally Convergent Method of Moving Asymptotes \cite{SVANBERG_2002} utilizing a maximum of 3 inner iterations per design iteration. Details of the optimization, post-processing, and evaluation processes are given in Appendix A.

\section{Results}

This section details eight studies conducted to investigate various aspects of the Raman scattering enhancement (Raman enhancement) problem. Study-specific parameters are given in each of the following subsections, while a general set of parameters used across all studies may be found in Appendix B.

\subsection{Comparison to known solutions} \label{SEC:REFERENCE_DESIGNS}

To demonstrate the proposed approach, we design a silver nanostructure that enhances Raman scattering from an isolated Raman molecule (Fig.~\ref{FIG:MODEL_PROBLEMS}A) excited at $\lambda_1 = 532$ nm, assuming a negligible Raman shift ($\lambda_2 = \lambda_1)$; non-negligible shifts are considered in Sec.~\ref{SEC:DIFFERENT_OBJECTIVES} and Sec.~\ref{SEC:RAMAN_SHIFT}. For context, the achieved enhancement is compared to enhancements generated by two simple parameter-optimized references.

Regarding the reference geometries, it is well known that placing a Raman molecule between two carefully scaled circular metallic cylinders (a coupled-sphere antenna) significantly enhances the Raman scattering process \cite{Huang2007,Zhu2011,Rechberger2003}, while placing it between two identical metallic triangular structures (a bowtie antenna) further enhances the process \cite{Hao2004,Dodson2013,Kaniber2016,Yue2017,Zhang2015}.  Both these references are parameter-optimized to maximize their performance at the targeted wavelengths. For the coupled-sphere antenna, the radius of the two discs is optimized over $R_{\mathrm{csa}} \in [10 \ \mathrm{ nm}, 50 \ \mathrm{ nm}]$, and we find the largest enhancement for $R_{\mathrm{csa}} = 48$ nm. The bowtie antenna is optimized by sweeping the tip-angle over, $\theta_{\mathrm{bta}} \in [10^{\mathrm{o}}, 180^{\mathrm{o}}]$, and the side length over $L_{\mathrm{bta}} \in [20 \ \mathrm{ nm}, 100 \ \mathrm{ nm}]$, and we find $\theta_{\mathrm{bta}} = 15^{\mathrm{o}}$ and $L_{\mathrm{bta}} = 70$ nm to yield the largest enhancement at $\lambda_1 = \lambda_2 = 532$ nm.

\begin{figure}[h!]
	\centering\includegraphics[width=0.475\linewidth]{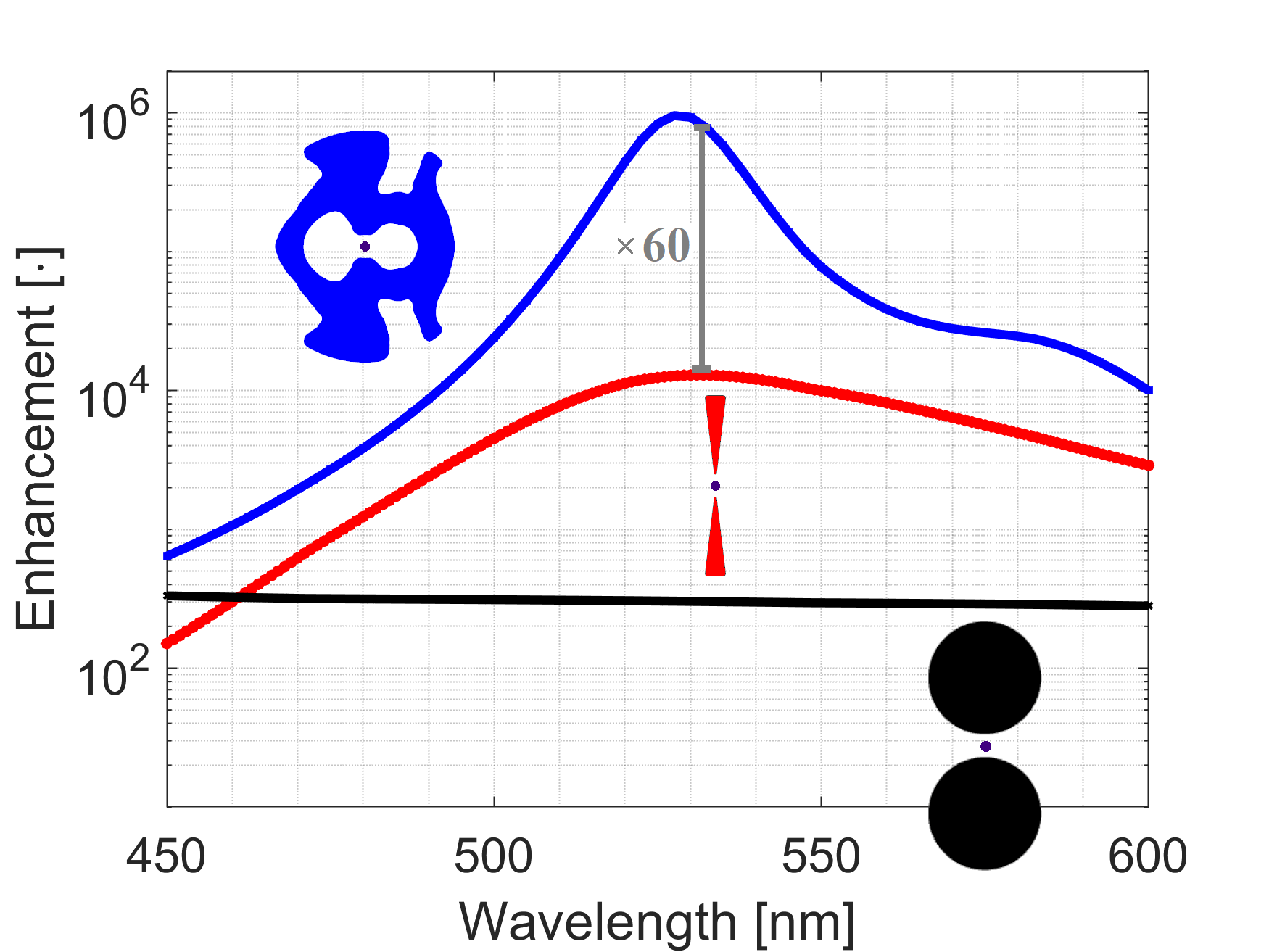}
	\caption{Raman enhancement as a function of wavelength, for a molecule placed at the center of different silver nanostructures (dark blue dot) relative to a molecule placed in free space. A topology-optimized structure (blue), a bowtie antenna (red) and a coupled-sphere antenna (black) are considered, all designed to maximize enhancement at $\lambda = 532$ nm. \label{FIG:REFERENCE_COMPARISON}}
\end{figure}

Figure \ref{FIG:REFERENCE_COMPARISON} shows the enhancement of the emitted power $\big($Eq.~(\ref{EQN:DESIGN})$\big)$ versus wavelength for each of the three structures, relative to the power emitted from the Raman molecule in free space. The coupled-sphere antenna (black) is seen to achieve a nearly wavelength-independent enhancement of $\approx3 \cdot 10^2$, while the bowtie antenna (red) achieves a peak enhancement of $\approx1.3 \cdot 10^4$ at $532$ nm. Interestingly, the intricate geometry of the topology-optimized nanostructure (blue), fully encapsulating the Raman molecule, achieves an enhancement of $\approx8.0 \cdot 10^5$. This is an increase by a factor of $\sim60\times$ relative to the bowtie antenna. The TO structure is in some sense a fusion of different features, tailored to enhance either the focusing or emission process, as will be studied more closely in Sec.~\ref{SEC:DIFFERENT_OBJECTIVES}. To our knowledge, no metallic structure for Raman enhancement similar to the optimized structure has previously been proposed, nor is such a structure likely to arise from applying any traditional design rules or intuition. 

A finding worth noting for this study, is that starting from an initial material configuration of  $\xi(\textbf{r}) = 0.0 \ \forall \textbf{r} \in \boldmath\Omega_{\mathrm{D}}$ results in the design process converging to a structure not encapsulating the Raman molecule. Importantly, this non-encapsulating design achieves an enhancement which is lower by more than a factor of $\sim2\times$ compared to the fully encapsulating design in Fig.~\ref{FIG:REFERENCE_COMPARISON}. In general, any structure found with gradient-based TO may only be a local optimum to the design problem. Alternative global optimization formulations are rarely practical or even feasible in large design spaces for non-convex problems \cite{SIGMUND_2011} and do anyway not guaranty global minima. Gradient-based optimization from different starting points (a "multistart" algorithm) can explore different local optima but in this work, the main goal is to find a structure much better than what could be easily designed by hand. Comparison to theoretical upper bounds is another route to gauging global optimality of TO structures \cite{Miller2016,Angeris2018}.

\subsection{Raman vs. focusing vs. emission} \label{SEC:DIFFERENT_OBJECTIVES}

A question worth asking, as it could potentially reduce the computational effort associated with the design procedure by a factor of two, is whether it is necessary to consider the full two-step Raman scattering process in the design procedure, or if it is possible to achieve similar performance by only considering an energy-focusing $\big($Eq.~(\ref{EQN:MAXWELL_PROBLEM1})$\big)$ or a dipole-emission  $\big($Eq.~(\ref{EQN:MAXWELL_PROBLEM2})$\big)$ problem. To answer this question using the proposed approach, we consider the following three cases. A dipole-emission case, an energy-focusing case and a case considering the full two-step process. All cases assume a 50 nm Raman shift, with $\lambda_1 = 532$ nm and $\lambda_2 = 582$ nm. 

For the dipole-emission case, the optimization problem $\big($Eqs.~(\ref{EQN:OBJECTIVE_FUNCTION})--(\ref{EQN:DESIGN_CON_1})$\big)$ remains unchanged, while the dipole considered in Eq.~(\ref{EQN:MAXWELL_PROBLEM2}) has its orientation fixed along the $y$ axis and its magnitude kept constant throughout the optimization, i.e. $\textbf{f}_2 = - \omega_2^2 \textbf{I} \langle1,0\rangle \delta(\textbf{r}-\textbf{r}_0)$, effectively removing the first model problem $\big($Eq.~(\ref{EQN:MAXWELL_PROBLEM1})$\big)$ as  $\textbf{f}_2 $ no longer depends on $\textbf{E}_1$. This corresponds to maximizing the local density of states for the dipole \cite{LIANG_AND_JOHNSON_2013}. For the energy-focusing case, the optimization problem is changed to one of solely maximizing $\vert \textbf{E}_1 \vert^2$ for an incident planewave at the position of the dipole \cite{WADBRO_ET_AL_2015,CHRISTIANSEN_VESTER-PETERSEN_2019} effectively removing the second model problem $\big($Eq.~(\ref{EQN:MAXWELL_PROBLEM2})$\big)$, as the emission process is no longer considered.

\begin{figure}[h!]
	\centering\includegraphics[width=0.425\linewidth]{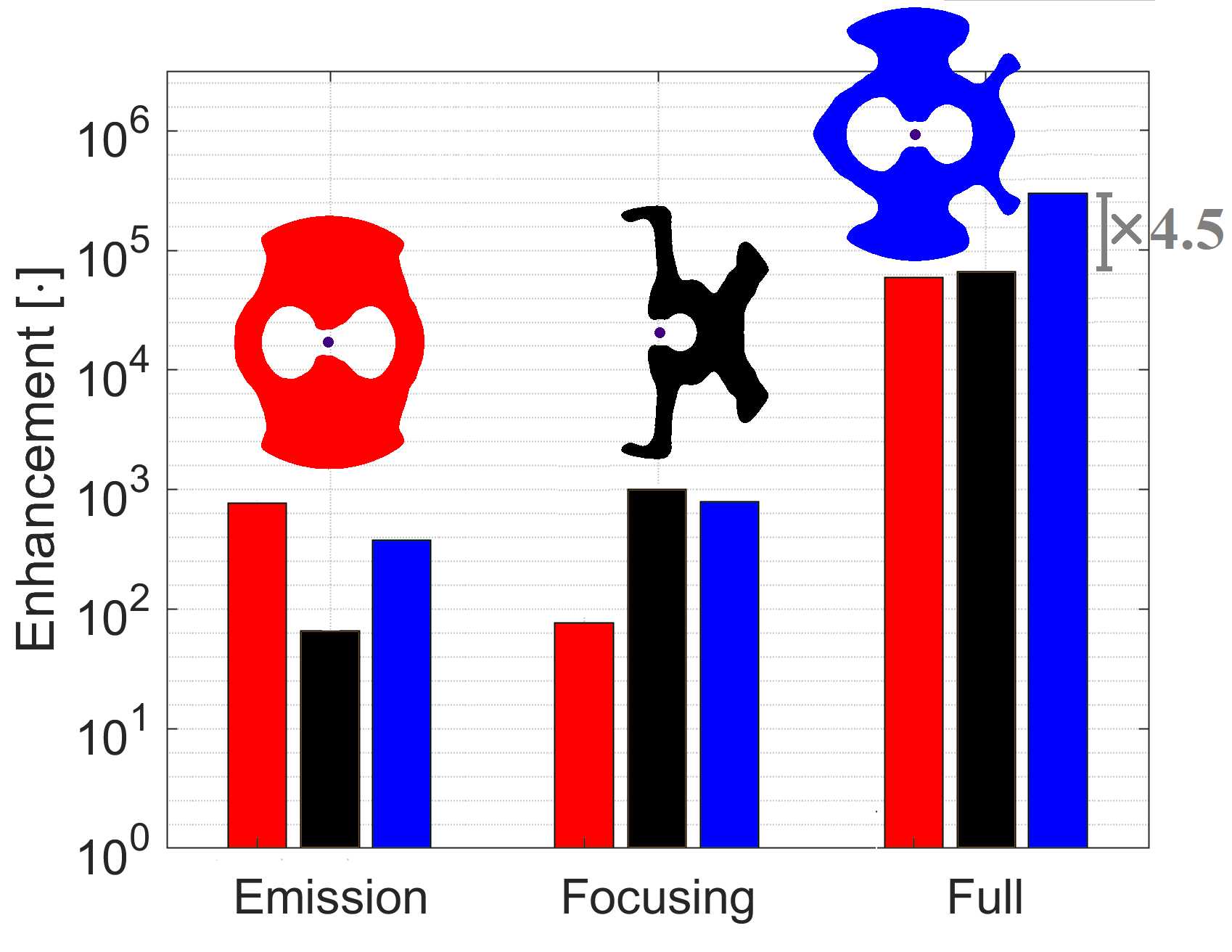}
	\caption{Objective enhancement achieved at $\lambda_1 = 532$, $\lambda_2 = 582$ nm by introducing optimized silver structures for pure dipole emission [Leftmost bar group]; pure energy focusing [Middle bar group]; full Raman scattering process [Rightmost bar group]. Structures optimized for dipole emission (red); energy focusing (black); full Raman scattering process (blue). The focal point (dipole position) is indicated using a dark blue dot.  \label{FIG:PROCESS_OPTIM}}
\end{figure}

Figure \ref{FIG:PROCESS_OPTIM} reports the enhancement of the three objectives attained by introducing the nanostructures optimized for the emission (red), focusing (black) and two-step Raman scattering (blue) process. The leftmost bar group reports the emission enhancement for a dipole with unit magnitude oriented along the $y$ axis, when placed at the center of each of the three designs. The middle bar group reports the enhancement of $\vert \textbf{E}_1 \vert^2$ at the dipole position for each design, while the rightmost bar group reports the enhancement achieved for the full two-step Raman scattering process. From the figure, it is seen that optimizing for a given process yields the best performance for that process. In particular, by explicitly optimizing for the two-step Raman scattering process, an increase in the enhancement by a factor of more than $4.5\times$ is achieved. It is interesting to observe that the design optimized for the full Raman scattering process qualitatively consists of a fusion of geometric features found in the focusing and the emission designs, e.g the closed cavity (emission) and the protruding features on the right side of the design (focusing). This suggests that a combination---and, by extension, a trade-off--- between the features is required to achieve the largest Raman enhancement.

\subsection{Size scaling}  \label{SEC:SIZE_SCALING}

It was recently shown that an upper bound for enhancing the Raman scattering process using metallic nanostructures scales linearly with the volume (area in 2d) of the design region \cite{MICHON_ET_AL_2019}. To investigate if this scaling is found using the proposed design approach, we perform a study considering three different outer radii of $\boldmath\Omega_{\mathrm{D}}$. These are $R_{\mathrm{out}} = 50$ nm (black), $R_{\mathrm{out}} = 75$ nm (red) and $R_{\mathrm{out}} = 100$ nm (blue), respectively.  All other parameters are kept constant across the three cases and an assumption of a negligible Raman shift is made, i.e. $\lambda_1 = \lambda_2 = 532$ nm.

\begin{figure}[h!]
	\centering\includegraphics[width=0.425\linewidth]{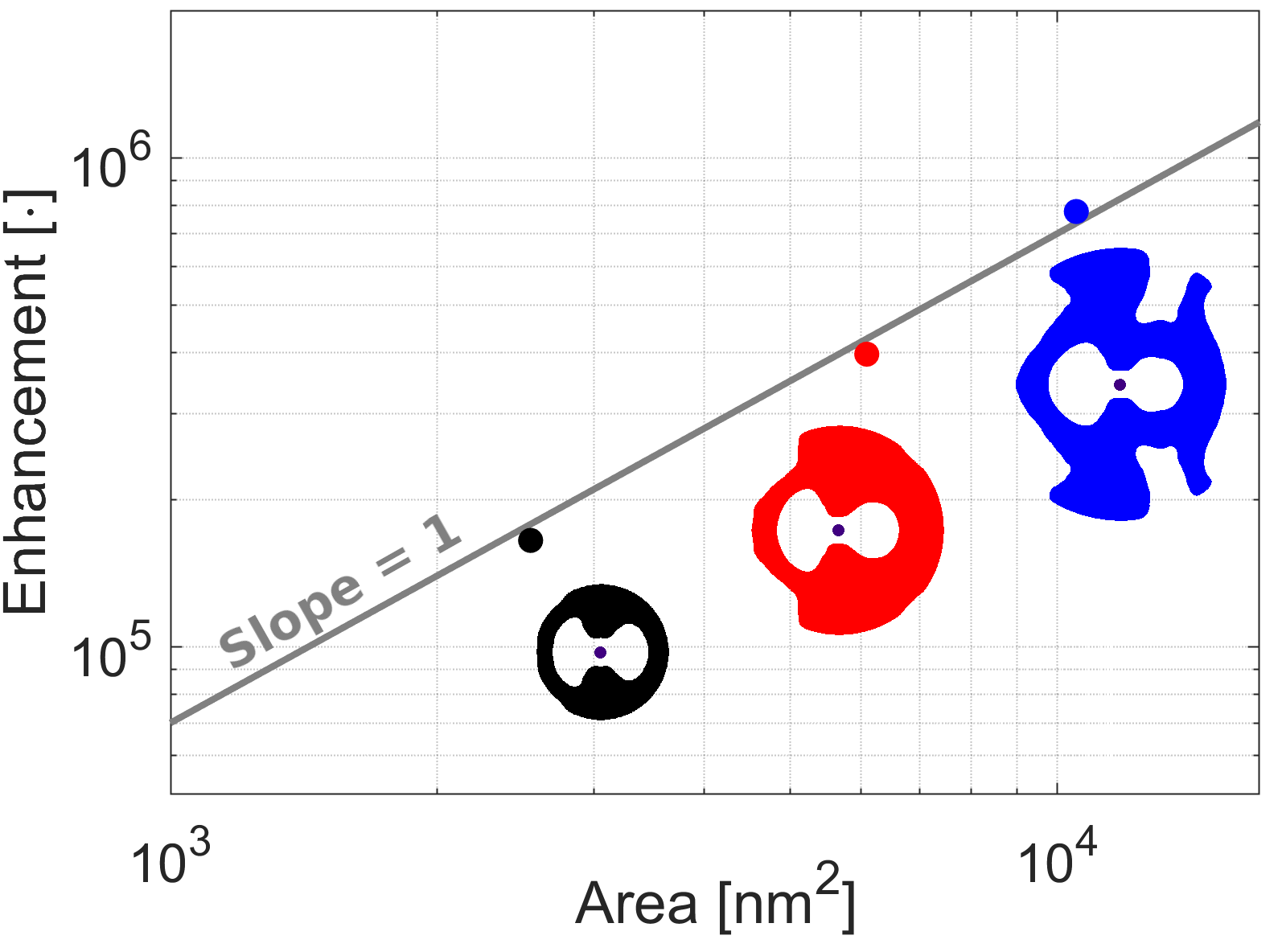}
	\caption{Raman enhancement versus $\boldmath\Omega_{\mathrm{D}}$ area, for a molecule placed at the center of the sliver structures (dark blue dot) relative to free space. A reference line (gray) with slope 1 is included to guide the eye. Nanostructures optimized for $R_{\mathrm{out}} = 50$ nm (black), $R_{\mathrm{out}} = 75$ nm (red) and $R_{\mathrm{out}} = 100$ nm (blue) are shown. Approximately linear scaling is observed. \label{FIG:SIZE_SCALING}}
\end{figure}

Figure \ref{FIG:SIZE_SCALING} shows the Raman enhancement relative to a molecule in free space versus the area of $\Omega_{\mathrm{D}}$ with the three designs and their performance plotted in black, red and blue and a trend line of slope 1 included to guide the eye. The data shows an approximately linear scaling of the emission enhancement with volume (area) in agreement with the upper bounds derived in \cite{MICHON_ET_AL_2019}, demonstrating that the proposed approach finds the expected volume scaling. 
 
\subsection{Material scaling}  \label{SEC:MATERIAL_SCALING}

The upper bound for enhancing the Raman scattering process, derived in \cite{MICHON_ET_AL_2019}, scales cubically with the material-related quantity $F = \vert \chi \vert^2 / \mathrm{Im}(\chi)$. Here a factor of $F^2$ stems from energy-focusing enhancement while the remaining factor of $F$ stems from dipole-emission enhancement. To investigate this behaviour, we conduct a materials study considering four metals, silver (Ag), gold (Au), copper (Cu) \cite{JOHNSON_AND_CHRISTY_1972}, and platinum (Pt) \cite{WERNER_2009}. All parameters, other than the material parameter $\varepsilon(\textbf{r})$, were kept constant across the four cases and a negligible Raman shift was assumed with $\lambda_1 = \lambda_2 = 532$ nm.

\begin{figure}[h!]
	\centering\includegraphics[width=0.425\linewidth]{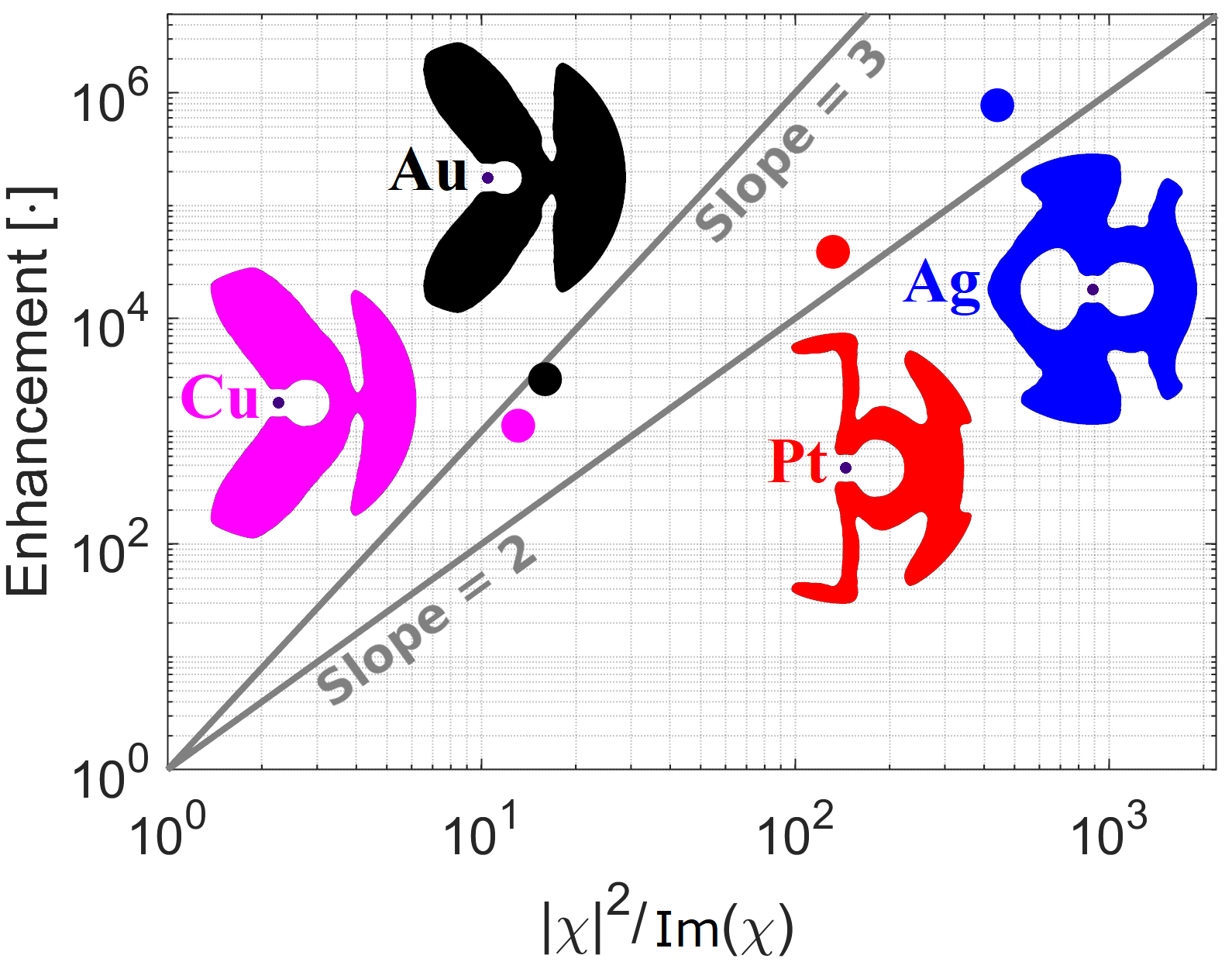}
	\caption{Raman enhancement for a Raman molecule (dark blue dot) versus $\vert \chi \vert^2 / \mathrm{Im}(\chi)$ relative to free space. Reference lines with slopes 2 and 3 (gray) are included. Optimized structures using Cu- (magenta), Au- (black), Pt- (red) and Ag-parameters (blue). \label{FIG:MATERIAL_SCALING}}
\end{figure}

Figure \ref{FIG:MATERIAL_SCALING} shows the Raman enhancement obtained by placing the Raman molecule at the center of the optimized structures relative to free space versus $\vert \chi \vert^2 / \mathrm{Im}(\chi)$ for the Cu- (magenta), Au- (black), Pt- (red) and Ag-design (blue). To guide the eye, two trend lines (gray) with slope 2 and slope 3 are plotted with the data. From a linear fit of the data, an approximate scaling with $F^2$ is observed. Trusting that TO indeed provides highly optimized results, as confirmed by the previous examples, this data suggests that it may not be possible to create a design which achieves ideal focusing- and ideal emission-enhancement simultaneously, but that a trade-off must be made. The observation is supported by the data in Fig.~\ref{FIG:PROCESS_OPTIM}, where significantly different geometries are observed when targeting the maximization of either focusing or emission, suggesting that different geometries are required to achieve either the highest attainable focusing or the highest attainable emission. Correspondingly, we find that the optimized structures (while vastly superior to bowtie antennas) still fall far short of the upper bound \cite{MICHON_ET_AL_2019}, especially for large values of $F$. Considering the analytical bound, future work adding an additional constraint coupling the two processes might lead to a tighter bound.

\subsection{Accounting for the Raman shift}  \label{SEC:RAMAN_SHIFT}

\begin{figure}[h!]
	\centering\includegraphics[width=0.4\linewidth]{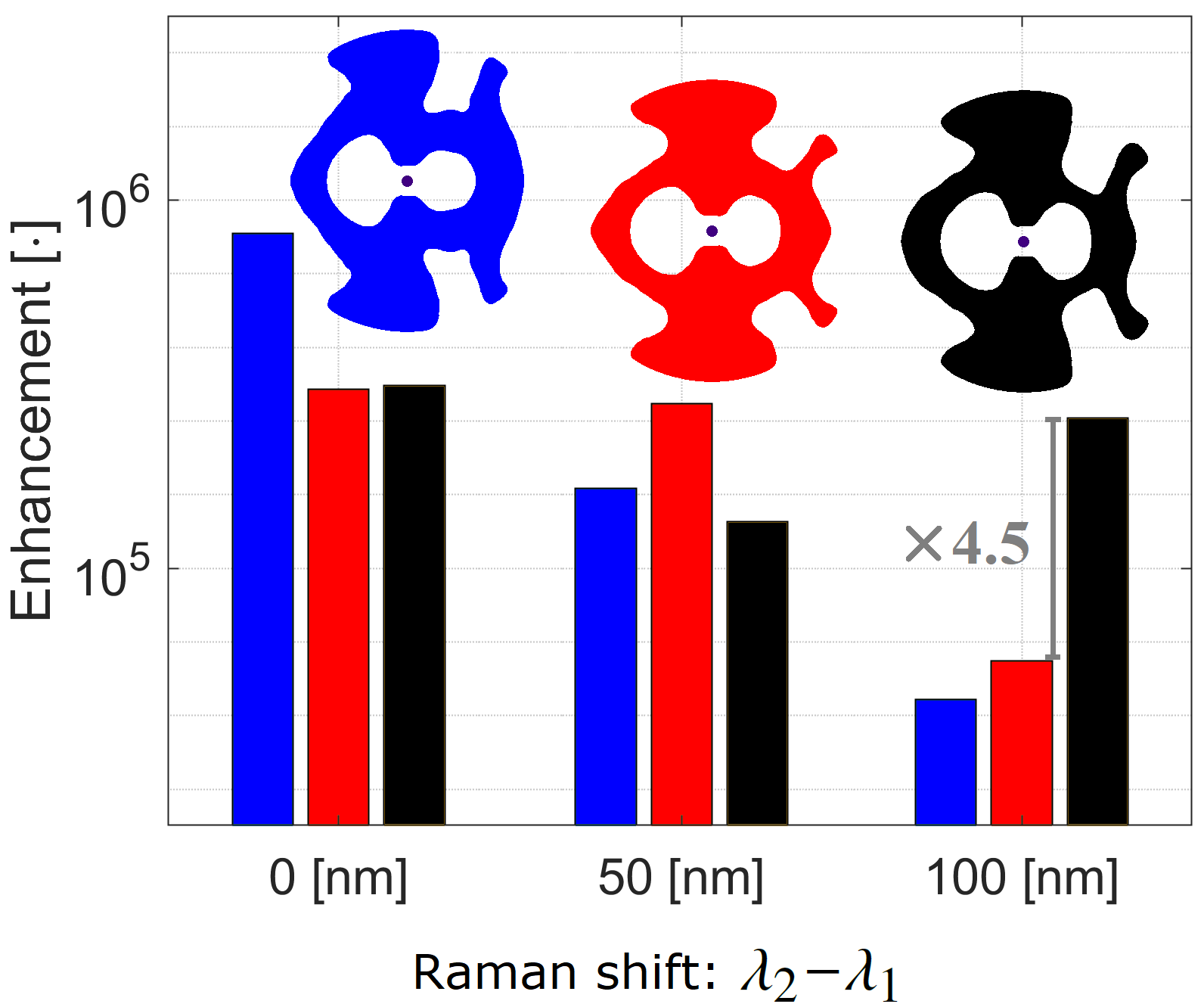}
	\caption{Raman enhancement for different Raman shifts for a Raman molecule (dark blue dot) placed at the centre of different silver nanostructures optimized for $\lambda_1 = 532$ nm and $\lambda_2 =$ 632 nm (black) $\lambda_2 =$ 582 nm (red) and $\lambda_2 =$ 532 nm (blue). The enhancement is reported for 0 nm [Leftmost], 50 nm [Middle] and 100 nm [Rightmost] bar groups. \label{FIG:RAMAN_SHIFT}}
\end{figure}

Depending on the molecule(s) and energy transitions of interest for a given Raman scattering problem, the Raman shift between the wavelengths $\lambda_1$ and $\lambda_2$ will be different. In this section, we investigate the benefits of accounting for the Raman shift in the design process. To this end, three cases are considered (Fig.~\ref{FIG:RAMAN_SHIFT}) with Raman shifts of 0 nm (blue), 50 nm (red), and 100 nm (black), respectively. The wavelength of the incident field is kept constant at $\lambda_1 = 532$ nm and the wavelength of the light emitted by the Raman molecule is adjusted accordingly. 

Figure \ref{FIG:RAMAN_SHIFT} presents the three designs along with the power-emission enhancement attained for each design when evaluated at 0 nm shift (leftmost bar group), 50 nm shift (middle bar group), and 100 nm shift (rightmost bar group). A first observation is that all three structures share the same overall geometrical features and thus in that sense look similar. However, looking more closely at each structure it is clear, that both the size and shape of each feature change across the designs. Furthermore, looking at the differences in enhancement, it is clear that these delicate feature changes are reflected in the performances of the structures. Unsurprisingly, each design exhibits the largest enhancement at the Raman shift it was optimized for. Specifically, an enhancement difference of a factor of $\sim1.6\times$ is observed for the 50 nm Raman shift while a factor of $\sim4.5\times$ is observed for the 100 nm Raman shift, clearly illustrating the performance benefit of accounting for the Raman shift as part of the design process.

\subsection{Wavelength dependence}  \label{SEC:WAVELENGTH_SCALING}

The importance of adjusting an electromagnetic structure to the operating wavelength, such as changing the length of an antenna, is well known. In this study we demonstrate that significantly larger design adjustments than simple scaling are required to achieve the best Raman enhancement for a given operating wavelength. It is shown how the optimized nanostructure geometry, as well as the emission enhancement, changes significantly with wavelength over the interval $\lambda_1 = \lambda_2 \in [250 \ \mathrm{ nm},490 \ \mathrm{ nm}]$. The study considers smaller designs with $R_{\mathrm{out}} = 30$ nm and assumes a negligible Raman shift.
In a sense, it probes a similar quantity as the material-dependence study in Sec.~\ref{SEC:MATERIAL_SCALING}, seeing as $\vert \chi \vert^2 / \mathrm{Im}(\chi)$ for silver varies by orders of magnitude across the wavelength interval. However, this study considers the added complexity that the wavelength changes along with $\varepsilon(\textbf{r})$.

\begin{figure}[h!]
	\centering\includegraphics[width=0.8\linewidth]{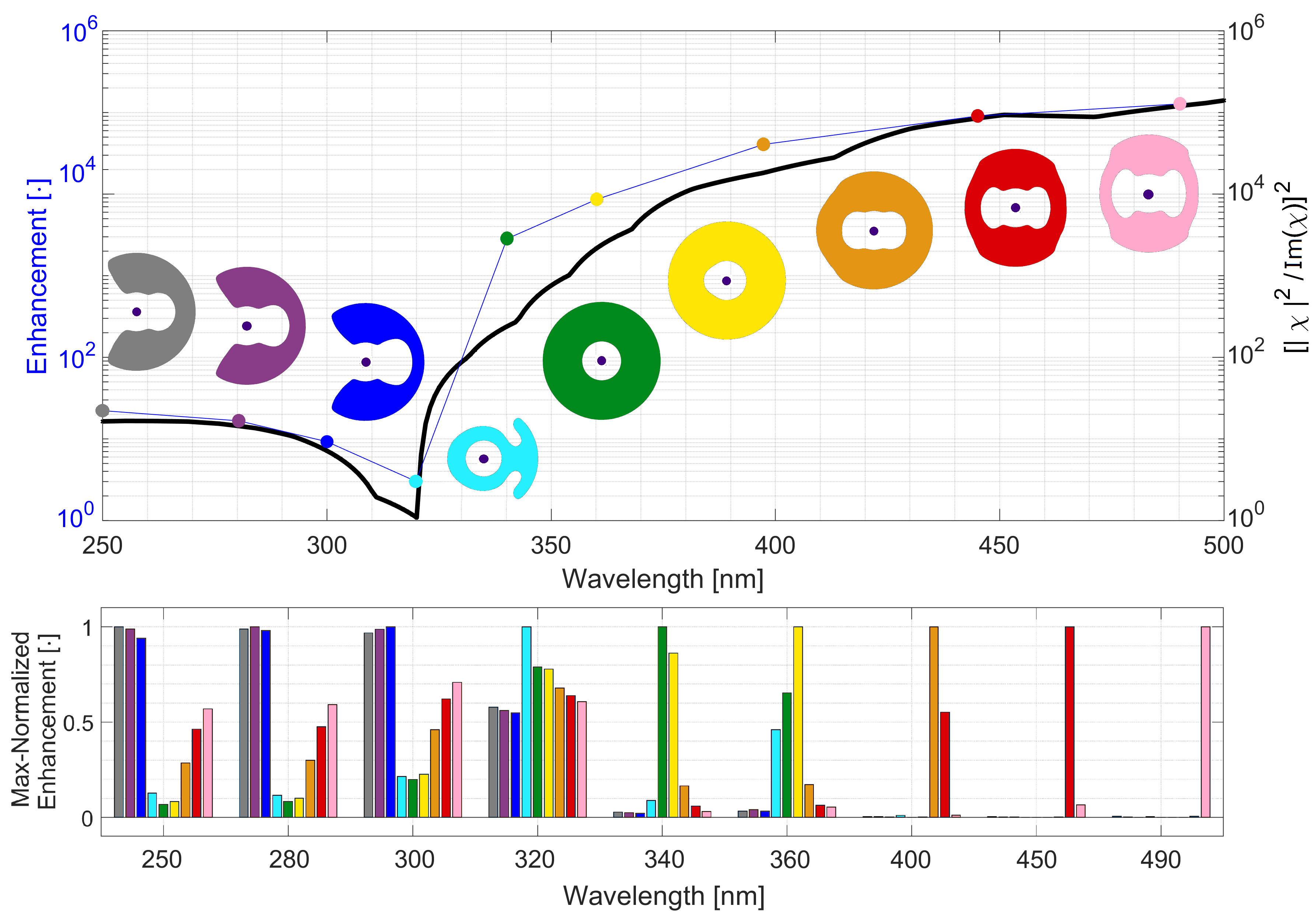}
	\caption{Raman enhancement relative to a molecule emitting in free space as a function of pump and emission wavelength ($\lambda_1 = \lambda_2$). [Top] Nine optimized silver nanostructures with outer design domain radius $R_{\mathrm{out}} =$  30 nm (rainbow) and the achieved enhancement (colored dots). The quantity $[\vert \chi \vert^2 / \mathrm{Im}(\chi)]^2$  is overlaid as a reference (black) and the position of the Raman molecule relative to the nanostructures is indicted using dark blue dots. [Bottom] The per-wavelength best design max-normalized enhancement for each of the nine wavelengths and each of the nine designs. \label{FIG:WAVELENGTH_SCALING}}
\end{figure}

The top panel of Fig.~\ref{FIG:WAVELENGTH_SCALING} shows nine silver nanostructures (rainbow-colored) optimized and evaluated for the reported wavelengths (colored dots). The quantity $[\vert \chi \vert^2 / \mathrm{Im}(\chi)]^2$ for silver is overlaid for easy reference (black). It is clearly observed that the enhancement factor of the optimized structures is strongly dependent on $[\vert \chi \vert^2 / \mathrm{Im}(\chi)]^2$. Furthermore, it is seen that the optimized nanostructure geometries change significantly from left to right, starting with a reflector-like geometry for $\lambda = 250$ nm and ending with a geometry resembling the red structure in Fig.~\ref{FIG:PROCESS_OPTIM} optimized to maximize dipole emission. Interestingly, around $\lambda = 320$ nm (near the minimum of $[\vert \chi \vert^2 / \mathrm{Im}(\chi)]^2$) the optimized structure is seen to experience a topological change which fundamentally changes the geometry from the reflector-like type to structures fully encapsulating the emitter. The bottom panel of Fig.~\ref{FIG:WAVELENGTH_SCALING} shows the per-wavelength max-normalized enhancement attained for each of the nine structures at each wavelength. From the panel it is seen that each of the optimized nanostructures outperformed the other structures at the wavelengths for which they were optimized. Furthermore, it is seen that as the wavelength increases the performance sensitivity increases. The data clearly shows the need for tailoring the nanostructure geometry to the particular operating conditions if the highest performance is sought.

\subsection{Out-of-plane Raman scattering}  \label{SEC:OUT_OF_PLANE_SCATTERING}

A common configuration for Raman-sensing applications is the distribution of Raman molecules over a surface, illuminated by some out-of-plane source with the scattered light collected out-of-plane as well. In the following study we demonstrate the strength of the proposed approach for such problem configurations (Fig.~\ref{FIG:MODEL_PROBLEMS}B).  

\begin{figure}[h!]
	\centering\includegraphics[width=0.8\linewidth]{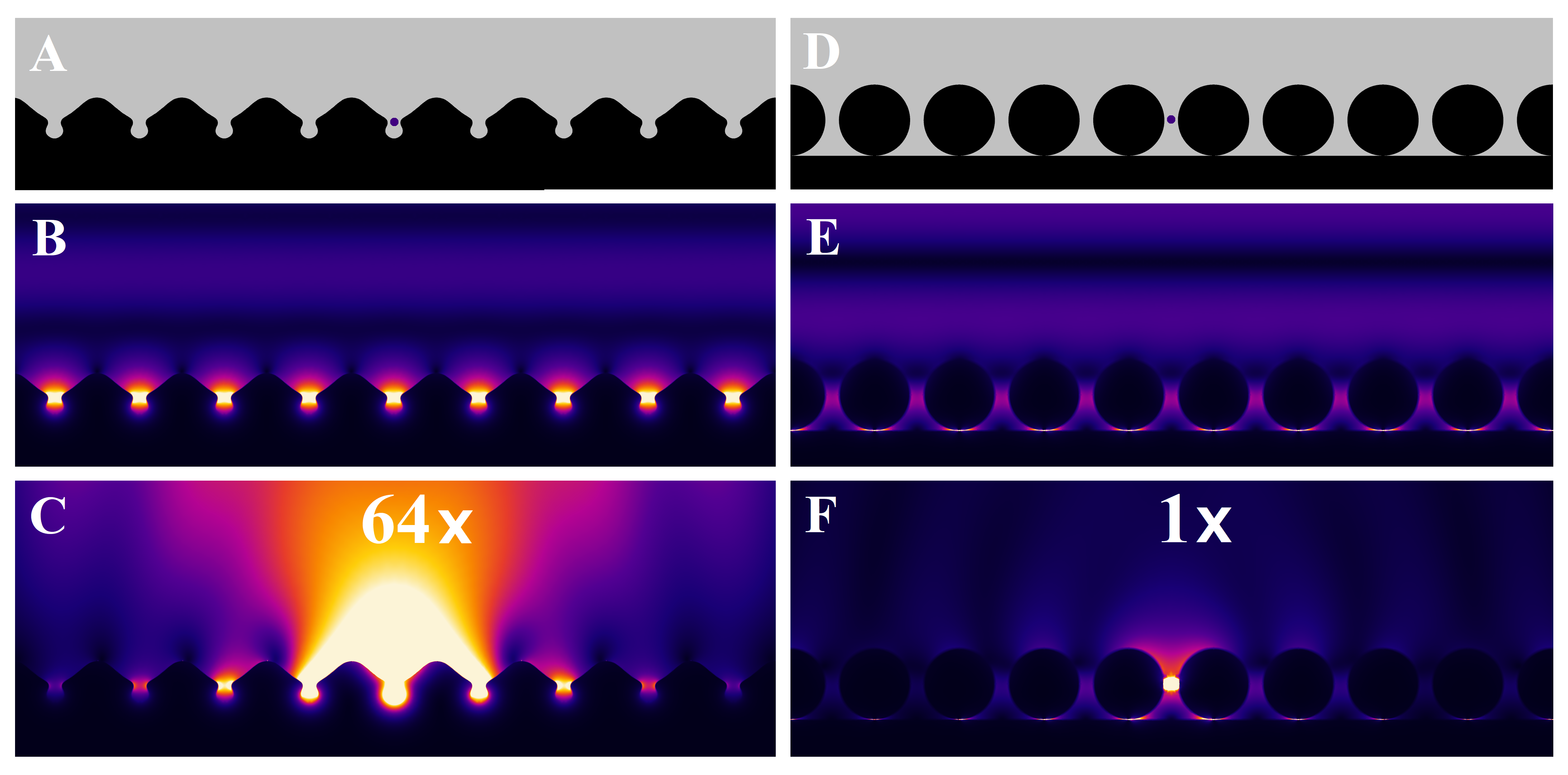}
	\caption{[left] Optimized and [right] reference periodically-structured platinum surfaces for the out-of-plane emission problem (Fig.~\ref{FIG:MODEL_PROBLEMS}B). \textbf{A)} Optimized surface (black) in water background (gray), with single Raman molecule (blue dot). \textbf{B)} $\vert \textbf{E}_1 \vert$-field at $\lambda_1= 532$ nm resulting from the interaction between the incident field and the topology-optimized surface. \textbf{C)} $\vert \textbf{E}_2 \vert$-field emitted at $\lambda_2= 549$ nm from a Raman molecule positioned at the center of the middle unit cell. \textbf{D)} Reference surface (black) in water background (gray), with single Raman molecule (blue dot). \textbf{E)} $\vert \textbf{E}_1 \vert$-field at $\lambda_1= 532$ nm  resulting from the interaction between the incident field and the reference surface. \textbf{F)} Emission $\vert \textbf{E}_2 \vert$-field emitted at $\lambda_2= 549$ nm from the Raman molecule. \label{FIG:OUT_OF_PLANE_RAMAN}}
\end{figure}

We consider a periodically-patterned platinum surface (period $a = 150$ nm) submerged in a water background ($\varepsilon_{\mathrm{H}_2\mathrm{O}} \approx 1.77$ around 500 nm). For computational simplicity, the study assumes a single Raman molecule per unit cell, situated at a fixed position at the center of the 150 nm x 200 nm design domain, which is again placed immediately on top of the platinum surface. A circular region of radius $R_{\mathrm{in}} = 10$ nm, centered at the Raman molecule, is kept free of platinum throughout the optimization. The model problem is excited at $\lambda_1 = 532$ nm with the Raman molecules emitting at $ \lambda_2 = 549$ nm. The molecules in the periodic array are assumed to scatter incoherently, as this most accurately models the physical scattering process. To account for this incoherent scattering, the array scanning method is used in the design process to compute the Raman scattering from a single molecule in the periodic model using 50 $k$-points in $[-\pi/a,\pi/a]$ \cite{SIGELMANN_ET_AL_1965,CAPOLINO_ET_AL_2007}. 

As a reference, we consider a periodic array ($a = 150$ nm) of circular platinum discs resting on top of the platinum surface, with the Raman molecules placed at the center of the gap between the discs (blue dot in Fig.~\ref{FIG:OUT_OF_PLANE_RAMAN}D). The separation distance from the molecule to the discs is taken to be 10 nm, identical to the radius of the platinum-free circular region in the optimization case.

Figure \ref{FIG:OUT_OF_PLANE_RAMAN} shows the optimized (left) and reference (right) surface structures and their behaviour under illumination at $\lambda_1 = 532$ nm with a single Raman molecule placed in the center unit cell. The top row shows the periodic platinum surface structures (black) in water background (gray) with the blue dot indicating the position of the Raman molecule. The middle row shows $\vert \textbf{E}_1 \vert$ at $\lambda_1 = 532$ nm for identical incident field strength using identical color scales. From the panels it is clear that the optimized structure provides a stronger focus of the incident field at the Raman molecule position. The bottom row presents $\vert \textbf{E}_2 \vert$ at $\lambda_2 = 549$ nm using identical color scales, clearly illustrating the significantly enhanced emission for the optimized surface structure. Computing the power emitted by the Raman molecule for both cases reveals a Raman enhancement by a factor of $\sim64\times$ for the optimized structure relative to the reference. While this example assumes a 2d model, it demonstrates the potential of applying the proposed approach to the design of nanostructures for full 3d out-of-plane Raman scattering problems.

\subsection{In-plane Raman scattering}  \label{SEC:IN_PLANE_SCATTERING}

In the final study we consider an in-plane Raman scattering problem with input and output waveguides (Fig.~\ref{FIG:MODEL_PROBLEMS}C). This study demonstrates the vast design freedom inherent to the proposed approach, which allows for the design of extreme enhancement structures in configurations where it would otherwise be difficult, if not impossible, to achieve good enhancement using conventional design techniques or intuition. 

\begin{figure}[h!]
	\centering\includegraphics[width=0.8\linewidth]{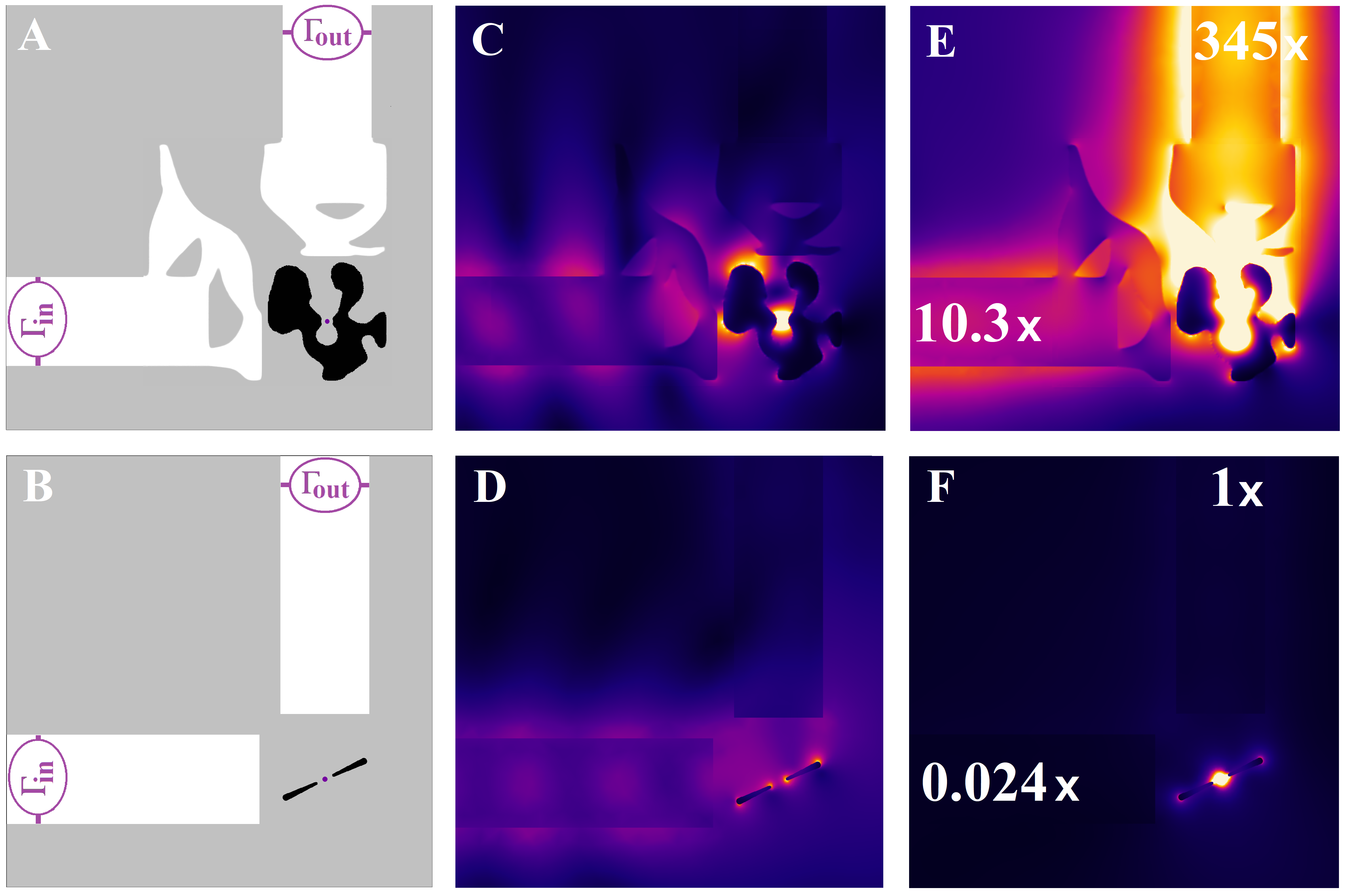}
	\caption{[Top row] Optimized and [bottom row] reference structures and fields for the in-plane emission problem (Fig.~\ref{FIG:MODEL_PROBLEMS}C). \textbf{A)} Optimized platinum (black) and Si$_3$N$_4$ structure (white) with Si$_3$N$_4$ input/output waveguides (white) and water background (gray). \textbf{B)} Reference platinum bowtie antenna (black) input/output waveguides (white) and water background (gray). \textbf{C}-\textbf{D)} $\vert \textbf{E}_1 \vert$-field at $\lambda_1= 532$ nm resulting from the interaction between the incident field through the lower left waveguide and the \textbf{C)} optimized and \textbf{D)} reference structure. \textbf{E}-\textbf{F)} $\vert \textbf{E}_2 \vert$-field emitted at $\lambda_2= 549$ nm from a Raman molecule positioned at the center of \textbf{E)} the optimized and \textbf{F)} reference platinum nanostructure. \label{FIG:IN_PLANE_RAMAN}}
\end{figure}

The objective of the design problem is to maximize the difference between the power emitted through $\Gamma_{\mathrm{out}}$ and through $\Gamma_{\mathrm{in}}$. This is achieved by designing a Pt nanostructure in $\boldmath\Omega_{\mathrm{D}_\mathrm{Metal}}$ together with a Si$_3$N$_4$ dielectric structure in $\boldmath\Omega_{\mathrm{D}_\mathrm{Dielectric}}$. Figure \ref{FIG:IN_PLANE_RAMAN}A shows the optimized nanostructure obtained by solving the design problem at $\lambda_1= 532$ nm,  $\lambda_2= 549$ nm.

As a reference, a bowtie antenna is parameter-optimized over its tip angle $\theta_{\mathrm{bta}} \in [10^{\mathrm{o}},90^{\mathrm{o}}]$, side length $L_{\mathrm{bta}} \in [20 \ \mathrm{ nm}, 100 \ \mathrm{ nm}]$, and rotation angle relative to horizontal $\theta_{\mathrm{rot}} \in [-90^{\mathrm{o}},90^{\mathrm{o}}]$. The parameters leading to the largest objective value are found to be $\theta_{\mathrm{bta}} = 10^{\mathrm{o}}$, $L_{\mathrm{bta}} = 62.5$ nm, and $\theta_{\mathrm{rot}} = 25^{\mathrm{o}}$. For the dielectric material distribution in the reference a simple extension of the two waveguides towards the bowtie antenna is used. The reference nanostructure is shown in Fig.~\ref{FIG:IN_PLANE_RAMAN}B.

Figure \ref{FIG:IN_PLANE_RAMAN}C--\ref{FIG:IN_PLANE_RAMAN}D shows the incident field at $\lambda_1= 532$ nm for the optimized and reference structure, respectively. Identical color scales are used for both plots, clearly showing the enhanced focusing of $\textbf{E}_1$ at the position of the Raman molecule. The rotated orientation of the reference structure, introduced to maximize the objective, means that the bowtie antenna is not capable of creating a highly localized focus between its tips. In contrast, the significantly more intricate geometry of the optimized design has no problem providing such enhancement.

Figure \ref{FIG:IN_PLANE_RAMAN}E--\ref{FIG:IN_PLANE_RAMAN}F shows the field emitted by the Raman molecule at $\lambda_2= 549$ nm for the optimized structure and the reference structure, respectively. Again, identical color scales are used for the two plots. From these it is obvious that the optimized structure generates a significantly enhanced emission relative to the reference. In fact, the difference between the two structures in terms of emitted power through $\Gamma_{\mathrm{out}}$ is a factor of $\sim345\times$, i.e. more than two orders of magnitude.

\section{Conclusion}  \label{SEC:CONCLUSION}

In this paper we proposed a TO-based approach for designing Raman scattering enhancing metallic nanostructures and presented a structure that increases the enhancement by a factor of $\sim 60\times$ compared to a parameter-optimized bowtie antenna (Sec.~\ref{SEC:REFERENCE_DESIGNS}).  Through a number of studies we documented: the importance of considering the full Raman scattering process in the design procedure (Sec.~\ref{SEC:DIFFERENT_OBJECTIVES}); that the expected linear scaling \cite{MICHON_ET_AL_2019} of the Raman enhancement with design volume (area) is achieved (Sec.~\ref{SEC:SIZE_SCALING}); that the Raman enhancement scales with $[\vert \chi \vert^2/\mathrm{Im}(\chi)]^2$ rather then the predicted scaling of $[\vert \chi \vert^2/\mathrm{Im}(\chi)]^3$ \cite{MICHON_ET_AL_2019}, possibly due to a trade off between the focusing and emission enhancement processes (Sec.~\ref{SEC:MATERIAL_SCALING}); the importance of accounting for the Raman shift in the design procedure (Sec.~\ref{SEC:RAMAN_SHIFT}); the importance of tailoring the nanostructure geometry to the operating wavelength as large geometric and associated performance changes occur as the wavelength is varied (Sec.~\ref{SEC:WAVELENGTH_SCALING}). Finally, we demonstrated that the TO-based approach may be used to achieve $\sim 64\times$ enhancement for out-of-plane Raman scattering (Sec.~\ref{SEC:OUT_OF_PLANE_SCATTERING}) and $\sim 345\times$ enhancement for in-plane Raman scattering in a two-waveguide setup (Sec.~\ref{SEC:IN_PLANE_SCATTERING}), relative to simple parameter optimized reference structures. 

While all studies in this work consider 2d model problems, the demonstration of extreme enhancements compared to well known reference geometries is promising. Hence, an important next step is the extension of the proposed approach to three dimensions. Applying the approach for Raman scattering problems modelling realistic operating conditions may help reveal hitherto undiscovered nanostructures for extreme Raman enhancement. If such structures are found, they may serve to improve existing Raman scattering based tools significantly as well as enable the development of novel tools.

An important topic for future work is to incorporate variability in the location of the Raman molecule. In some experimental situations, the Raman molecules are suspended in a fluid and the objective is to maximize the \emph{average} emission over all possible molecule locations \cite{LeRu2007,Ding2017,Lee2018}. Naively, this would require solving the model problem for a vast number of different dipole locations and averaging the emission, making such a task computationally infeasible for all but the smallest of problems. However, efficient "trace" formulations have been developed for related problems involving a distribution of random emitters---thermal emission \cite{Rodriguez2012,Polimeridis2015} and spontaneous emission \cite{Polimeridis2015}---and we believe that similar techniques are applicable to the Raman problem.

Another future step is the investigation of the sensitivity of the optimized structures towards perturbations of their geometry, such as those encountered in fabrication. Results in previous studies on designing plasmonic nanostructures using TO, e.g. \cite{VESTER-PETERSEN_2017}, as well as several results in this work, suggest that the optimized nanostructures are sensitive to geometric variation, as small variations lead to large performance changes. By employing well known robust optimization techniques \cite{WANG_ET_AL_2011,CHRISTIANSEN_ET_AL_2015} it may be possible to reduce the geometric sensitivity of the nanostructures, hereby bridging a gap between simulations and experiment. 


\appendix
\section*{Appendix A. Optimization, post processing and evaluation procedure}  \label{APN:OPTIMIZATION_EVALUATION}

The physics is modelled in COMSOL Multiphysics \cite{COMSOL54} and the optimization problem is solved using the Globally Convergent Method of Moving Asymptotes (GCMMA) \cite{SVANBERG_2002}.  \\

\noindent The following stopping criterion is used to terminate the optimization: \\

\begin{algorithmic}
	\If {$i \geq i_{\mathrm{min}}$}
	\If {$\beta < \beta_{\mathrm{max}}$}
	\State $\beta = 2 \beta$
	\State $i = 1$
	\Else
	\If {$ \vert \Phi_{i} - \Phi_{i-n} \vert / \vert \Phi_{i}  \vert  \leq 0.001 \ \forall \ n  \  \lbrace  1,2,...,10 \rbrace $}  
	\State Terminate optimization.
	\EndIf
	\EndIf
	\EndIf \\
\end{algorithmic}

\noindent Here $i$ denotes the current optimization iteration, $i_{\mathrm{min}} = 70$ denotes the minimum number of iterations taken. $\beta$ denotes the thresholding strength used in the filtering procedure and $\beta_{\mathrm{max}} = 32$. $\Phi_{i}$ denotes the objective function value at the $i$'th iteration and $n \in \mathbb{N}^{+}$. \\

\noindent After the design process is completed a post-processing procedure is performed to extract the final nanostructure from the optimized material distribution. In this procedure the final $\epsilon(\textbf{r})$-field is sampled in a Cartesian (x,y)-grid with 0.1 nm resolution. The sampled distribution is smoothed using a simple cone-shaped kernel \cite{BOURDIN_ET_AL_2001} with 1.5 nm filter radius to remove all sub-nanometer kinks left by the 1 nm topology optimization mesh. The final geometry is then extracted as the $0.5$-contour of the smoothed field. Finally, the geometry is rescaled to have an inner radius of $r_{\mathrm{inner}} = 10$ nm for consistency across designs. 

The reported performances and fields are evaluated using the final post processed design, discretized using an unstructured triangular body fitted mesh with 1 nm side length for the structure. Quadratic continuous Lagrange basis functions are used to resolve the physics during the evaluation step. 

Note that the exact position and magnitude of the Raman enhancement peak depend strongly on the final geometry. This means that a size scaling of a structure by a few percent may result in a shift of several nm in the emission peak position as well as a change in the overall enhancement. 

\section*{Appendix B. Study parameters} \label{APN:STUDY_PARAMETERS}

This appendix lists the parameters used in setting up the model and associated optimization problems. Unless specified otherwise below or in the individual sections detailing each study, the parameters listed in the following are used.

\subsection*{B. 1. Model Domain}

For the model domain sketched in Fig.~\ref{FIG:MODEL_PROBLEMS}A the following values are used: \\

\noindent The model domain $\boldmath{\Omega}$ is a square with side length, $L_{\boldmath{\Omega}} = 600$ nm.  $\boldmath{\Omega}$ is surrounded by a perfectly matched layer  with a depth of $D_{\mathrm{PML}}= 300$ nm. The design domain $\boldmath{\Omega}_{\mathrm{D}}$ is centered in  $\boldmath{\Omega}$ and taken to be a perfect 2d torus with inner radius, $R_{\mathrm{in}} = 10$ nm and outer radius, $R_{\mathrm{out}} = 100$ nm. The Raman molecule is modelled as a point dipole source placed at the center of $\boldmath{\Omega}_{\mathrm{D}}$. The power emitted by the dipole at $\lambda_2$ is computed by evaluating Eq.~(\ref{EQN:DESIGN}) over $\boldmath{\Gamma}_{\mathrm{RE}},$ a circular curve centered on the dipole with radius: $R_{\boldmath{\Gamma}_{\mathrm{RE}]}} = 250$ nm. \\

\noindent For the model domain sketched in Fig.~\ref{FIG:MODEL_PROBLEMS}B the following values are used: \\

\noindent  The model domain $\boldmath{\Omega}$ is a rectangle of width, $\mathrm{a} = W_{\boldmath{\Omega}} = 150$ nm and height $H_{\boldmath{\Omega}} = 600$ nm. $\boldmath{\Omega}$ is truncated from above by perfectly matched layer  with a depth of $D_{\mathrm{PML}}= 300$ nm.  $\boldmath{\Omega}$ is truncated from below using a perfect electric conductor boundary condition: $\textbf{n} \times \textbf{E} = 0$. $\boldmath{\Omega}$ is truncated by a Floquet Bloch boundary condition on the left and right boundaries to model the $x$-periodicity. The top 500 nm of $\boldmath{\Omega}$ is taken to contain water ($\varepsilon_r \approx 1.77$). The bottom 100 nm of $\boldmath{\Omega}$ is taken to contain platinum \cite{WERNER_2009}. The design domain $\boldmath{\Omega}_{\mathrm{D}}$ is placed immediately on top of the platinum region. The design domain $\boldmath{\Omega}_{\mathrm{D}}$ has a width of $W_{\boldmath{\Omega}_{\mathrm{D}}} = 150$ nm and a height $H_{\boldmath{\Omega}_{\mathrm{D}}} = 200$ nm. The design domain $\boldmath{\Omega}_{\mathrm{D}}$ has a region fixed to contain water at its center with radius $R_{\mathrm{in}} = 10$ nm. The Raman molecule is modelled as a point dipole source placed at the center of $\boldmath{\Omega}_{\mathrm{D}}$. The power emitted by the dipole at $\lambda_2$ is computed by evaluating Eq.~(\ref{EQN:DESIGN}) over $\boldmath{\Gamma}_{\mathrm{RE}},$ a straight horizontal line 50 nm below the top boundary of $\boldmath{\Omega}$.\\

\noindent For the model domain sketched in Fig.~\ref{FIG:MODEL_PROBLEMS}C the following values are used: \\

\noindent The model domain $\boldmath{\Omega}$ is a square with side length, $L_{\boldmath{\Omega}} = 1200$ nm centered at (0 nm,0 nm). $\boldmath{\Omega}$ is surrounded by a perfectly matched layer  with a depth of $D_{\mathrm{PML}}= 300$ nm. The design domain $\boldmath{\Omega}_{\mathrm{D}_{\mathrm{Metal}}}$ is  centered at (-300 nm,-300 nm).  $\boldmath{\Omega}_{\mathrm{D}_{\mathrm{Metal}}}$ is a square of side length $L_{\boldmath{\Omega}_{\mathrm{D}_{\mathrm{Metal}}}} = 200$ nm with a circular hole of radius $R_{\mathrm{in}} = 10$ nm at its center. The Raman molecule is modelled as a point dipole source placed at the center of $\boldmath{\Omega}_{\mathrm{D}}$. The design domain $\boldmath{\Omega}_{\mathrm{D}_{\mathrm{Dielectric}}}$ is  centered at (-200 nm,-200 nm).  $\boldmath{\Omega}_{\mathrm{D}_{\mathrm{Dielectric}}}$ is the difference between a square of side length $L_{\boldmath{\Omega}_{\mathrm{D}_{\mathrm{Dielectric}}}} = 400$ nm and $\boldmath{\Omega}_{\mathrm{D}_{\mathrm{Metal}}}$. The input waveguide runs horizontally and is centered at $y = -300$ nm. It starts at the left edge of $\boldmath{\Omega}$ and runs until $\boldmath{\Omega}_{\mathrm{D}_{\mathrm{Dielectric}}}$. The output waveguide runs vertically and is centered at $x = 300$ nm. It starts at the top edge of $\boldmath{\Omega}$ and runs until $\boldmath{\Omega}_{\mathrm{D}_{\mathrm{Dielectric}}}$. The width of both waveguides is 150 nm. The power emitted by the dipole at $\lambda_2$ into the input and output waveguides is computed by evaluating Eq.~(\ref{EQN:DESIGN}) over $\boldmath{\Gamma}_{\mathrm{in}}$ and $\boldmath{\Gamma}_{\mathrm{out}}$ respectively. $\boldmath{\Gamma}_{\mathrm{in}}$ is a vertical line through the input waveguide at $x = -300$ nm. $\boldmath{\Gamma}_{\mathrm{out}}$ is a horizontal line through the output waveguide at $t = 300$ nm. 

\subsection*{B. 2. Numerical Solution}
For the numerical solution of all model problems the geometry is discretized using an unstructured first order finite element mesh \cite{BOOK_FEM_EM_JIN}. The design domain is discretized using triangular elements with a uniform side length of $h = 1$ nm, dictating the resolution of the design. The remaining model domain is discretized using a non-uniform mesh of triangular elements with a smallest side lengths of $h=1$ nm near the design domain and a maximum side length of $h = 1/16$ effective wavelength ($\lambda / n$), to ensure the accuracy of the model. \\

\noindent For the model problems sketched in Fig.~\ref{FIG:MODEL_PROBLEMS}A-\ref{FIG:MODEL_PROBLEMS}B the incident-field problem $\big($Eq.~(\ref{EQN:MAXWELL_PROBLEM1})$\big)$, is solved using the scattered-field approach \cite{BOOK_FEM_EM_JIN}, where the background field is taken to be a perfect planewave. \\

\noindent For the model problems sketched in Fig.~\ref{FIG:MODEL_PROBLEMS}C the incident-field problem is solved using the total-field approach \cite{BOOK_FEM_EM_JIN}. First the lowest-order mode confined to the input waveguide is computed. This mode is then used to excite the system and a total-field problem is solved. \\

\noindent For all model problems the emitter problem $\big($Eq.~(\ref{EQN:MAXWELL_PROBLEM2})$\big)$ is solved for the total field.

\subsection*{B. 3. Physics Related}
The physics related parameters are chosen as follows:  \\

\noindent   The wavelength of the incident field is taken to be $\lambda_1 = 532$ nm. The wavelength of the emitted field is taken to be $\lambda_2 = 532$ nm. The relative permittivity and relative permeability of air are taken to be $\varepsilon_{\mathrm{air}} = \mu_{\mathrm{air}} = 1.0$. The relative permittivity and relative permeability of water are taken to be $\varepsilon_{\mathrm{water}} = 1.77$, $\mu_{\mathrm{water}} = 1.0$ at the operating wavelengths. The relative permittivity and relative permeability of silver (Ag), gold (Au) and copper (Cu) are taken from \cite{JOHNSON_AND_CHRISTY_1972}. The relative permittivity and relative permeability of platinum (Pt) are taken from \cite{WERNER_2009}. The relative permittivity and relative permeability of Si$_3$N$_4$ are taken to be $\varepsilon_{\mathrm{Si}_3\mathrm{N}_4} = 2.05$, $\mu_{\mathrm{water}} = 1.0$ at the operating wavelengths. The speed of light is taken to be $c = 3 \cdot 10^8$ m/s.

\subsection*{B. 4. Design Related}

For the designs considering the model problem sketched in Fig.~\ref{FIG:MODEL_PROBLEMS}A mirror symmetry is imposed on the designs normal to the horizontal axis due to the nature of the model problem, where the planewave in Eq.~(\ref{EQN:MAXWELL_PROBLEM1}) propagates through the domain along the horizontal axis. \\

\noindent For the designs considering the model problem sketched in Fig.~\ref{FIG:MODEL_PROBLEMS}B mirror symmetry is imposed on the design normal to the vertical axis due to the nature of the model problem, where the planewave in Eq.~(\ref{EQN:MAXWELL_PROBLEM1}) propagates through the domain along the vertical axis.

\subsection*{B. 5. Optimization Related}

For the model problem sketched in Fig.~\ref{FIG:MODEL_PROBLEMS}A the following values are used: \\

\noindent As an initial guess for all optimizations a full metal disc is used, i.e. $\xi_{\mathrm{initial}}(\textbf{r}) = 1 \ \forall \ \textbf{r} \in \boldmath{\Omega}_{\mathrm{D}}$.  The filter radius in the smoothing operation is, $r_f = 10$ nm. ($r_f = 5$ nm was used for the study detailed in Sec.~\ref{SEC:WAVELENGTH_SCALING}.) The thresholding level in the thresholding operation is, $\eta = 0.5$. The initial thresholding strength is, $\beta_{\mathrm{initial}} = 8$. The thresholding strength is increased gradually during the optimization through the values, $\beta = 8,16,32$. A minimum length scale of all features in the designs is ensured using geometric length-scale constraints with $c_{\mathrm{LS}} = 625$, $\eta_{e} = 0.75$, $\eta_{d} = 0.25$. The length-scale constraints are only enforced for $\beta = 32$ to allow the design to form freely without limiting topology changes in the earlier stages of the optimization.  \\

\noindent For the model problem sketched in Fig.~\ref{FIG:MODEL_PROBLEMS}B the following values are used:  \\

\noindent As an initial guess a full metal region is used, i.e. $\xi_{\mathrm{initial}}(\textbf{r}) = 1 \ \forall \ \textbf{r} \in \boldmath{\Omega}_{\mathrm{D}}$. The filtering radius used in the smoothing operation is, $r_f = 10$ nm. The thresholding level used is in the thresholding operation is, $\eta = 0.5$. The initial thresholding strength used in the thresholding operation is, $\beta_{\mathrm{initial}} = 8$. The thresholding strength is increased gradually during the optimization through the values, $\beta = 8,16,32$. A minimum length scale of all features in the designs is ensured using geometric length-scale constraints using $c_{\mathrm{LS}} = 625$, $\eta_{e} = 0.75$, $\eta_{d} = 0.25$. The length-scale constraints are only enforced for $\beta = 32$ to allow the design to form freely without limiting topology changes in the earlier stages of the optimization.  \\

\noindent For the model problem sketched in Fig.~\ref{FIG:MODEL_PROBLEMS}C the following values are used:  \\

\noindent As an initial guess for $\boldmath{\Omega}_{\mathrm{D}_{\mathrm{Metal}}}$  a full metal region is used, i.e. $\xi_{\mathrm{initial}}(\textbf{r}) = 1 \ \forall \ \textbf{r} \in \boldmath{\Omega}_{\mathrm{D}_{\mathrm{Metal}}}$. As an initial guess for $\boldmath{\Omega}_{\mathrm{D}_{\mathrm{Dielectric}}}$ a full dielectric region is used, i.e. $\xi_{\mathrm{initial}}(\textbf{r}) = 1 \ \forall \ \textbf{r} \in \boldmath{\Omega}_{\mathrm{D}_{\mathrm{Dielectric}}}$. The filtering radius in the smoothing operation is: $r_f = 10$ nm. The thresholding level in the thresholding operation is: $\eta = 0.5$. The initial thresholding strength used is in the thresholding operation is: $\beta_{\mathrm{initial}} = 8$. The thresholding strength is increased gradually during the optimization through the values: $\beta = 8,16,32$. A minimum length scale of all features in the designs is ensured using geometric length-scale constraints using $c_{\mathrm{LS}} = 625$, $\eta_{e} = 0.75$, $\eta_{d} = 0.25$. The length-scale constraints is only imposed for $\beta = 32$ to allow the design to form freely without limiting topology changes in the earlier stages of the optimization.

\section*{Funding} 

NATEC (NAnophotonics for TErabit Communications) Centre (8692) awarded by Villum Fonden; U.S. Army Research Office (W911NF-13-D-0001); National Science Foundation (1453218,1709212). 

\section*{Acknowledgements} 

The authors thank Juejun Hu for insightful discussions. 
 
\section*{Disclosures} 

The authors declare no conflicts of interest. 

\bibliographystyle{ieeetr} 
\bibliography{References}   

\begin{thebibliography}{10}

\bibitem{BOOK_TOPOPT_BENDSOE}
M.~P. Bendsøe and O.~Sigmund, {\em Topology Optimization}.
\newblock Springer, 2003.

\bibitem{JENSEN_SIGMUND_2011}
J.~S. Jensen and O.~Sigmund, ``Topology optimization for nano-photonics,'' {\em
  Laser \& Photonics Reviews}, vol.~5, pp.~308--321, 2011.

\bibitem{MOLESKY_2018}
S.~Molesky, Z.~Lin, A.~Y. Piggott, W.~Jin, J.~Vuckovic, and A.~W. Rodriguez,
  ``Inverse design in nanophotonics,'' {\em Nature Photonics}, vol.~12,
  pp.~659--670, 2018.

\bibitem{Long1977}
D.~A. Long, {\em {Raman spectroscopy}}.
\newblock McGraw-Hill, 1977.

\bibitem{Turrell1996}
G.~Turrell and J.~Corset, {\em {Raman Microscopy: Developments and
  Applications}}.
\newblock Academic Press, 1996.

\bibitem{Colthup1990}
N.~B. Colthup, L.~H. Daly, and S.~E. Wiberley, {\em {Introduction to Infrared
  and Raman Spectroscopy}}.
\newblock Academic Press, 1990.

\bibitem{Huang2007}
W.~Huang, W.~Qian, P.~K. Jain, and M.~A. El-Sayed, ``{The effect of plasmon
  field on the coherent lattice phonon oscillation in electron-beam fabricated
  gold nanoparticle pairs},'' {\em Nano Letters}, vol.~7, no.~10,
  pp.~3227--3234, 2007.

\bibitem{Zhu2011}
W.~Zhu, M.~G. Banaee, D.~Wang, Y.~Chu, and K.~B. Crozier, ``{Lithographically
  fabricated optical antennas with gaps well below 10 nm},'' {\em Small},
  vol.~7, no.~13, pp.~1761--1766, 2011.

\bibitem{Rechberger2003}
W.~Rechberger, A.~Hohenau, A.~Leitner, J.~R. Krenn, B.~Lamprecht, and F.~R.
  Aussenegg, ``{Optical properties of two interacting gold nanoparticles},''
  {\em Optics Communications}, vol.~220, no.~1-3, pp.~137--141, 2003.

\bibitem{Hao2004}
E.~Hao and G.~C. Schatz, ``{Electromagnetic fields around silver nanoparticles
  and dimers},'' {\em Journal of Chemical Physics}, vol.~120, no.~1,
  pp.~357--366, 2004.

\bibitem{Dodson2013}
S.~Dodson, M.~Haggui, R.~Bachelot, J.~Plain, S.~Li, and Q.~Xiong, ``{Optimizing
  electromagnetic hotspots in plasmonic bowtie nanoantennae},'' {\em Journal of
  Physical Chemistry Letters}, vol.~4, no.~3, pp.~496--501, 2013.

\bibitem{Kaniber2016}
M.~Kaniber, K.~Schraml, A.~Regler, J.~Bartl, G.~Glashagen, F.~Flassig,
  J.~Wierzbowski, and J.~J. Finley, ``{Surface plasmon resonance spectroscopy
  of single bowtie nano-antennas using a differential reflectivity method},''
  {\em Scientific Reports}, vol.~6, no.~1, p.~23203, 2016.

\bibitem{Yue2017}
W.~Yue, Z.~Wang, J.~Whittaker, F.~Lopez-Royo, Y.~Yang, and A.~V. Zayats,
  ``{Amplification of surface-enhanced Raman scattering due to
  substrate-mediated localized surface plasmons in gold nanodimers},'' {\em
  Journal of Materials Chemistry C}, vol.~5, no.~16, pp.~4075--4084, 2017.

\bibitem{Zhang2015}
J.~Zhang, M.~Irannejad, and B.~Cui, ``{Bowtie Nanoantenna with Single-Digit
  Nanometer Gap for Surface-Enhanced Raman Scattering (SERS)},'' {\em
  Plasmonics}, vol.~10, no.~4, pp.~831--837, 2015.

\bibitem{Moskovits1985}
M.~Moskovits, ``Surface-enhanced spectroscopy,'' {\em Reviews of Modern
  Physics}, vol.~57, pp.~783--826, 1985.

\bibitem{Campion1998}
A.~Campion and P.~Kambhampati, ``{Surface-enhanced Raman scattering},'' {\em
  Chemical Society Reviews}, vol.~27, no.~4, pp.~241--250, 1998.

\bibitem{Kneipp2006}
K.~Kneipp, M.~Moskovits, and H.~Kneipp, {\em {Surface-Enhanced Raman
  Scattering}}.
\newblock Springer, 2006.

\bibitem{leru2008}
E.~{Le Ru} and P.~Etchegoin, {\em Principles of Surface-Enhanced Raman
  Spectroscopy: and Related Plasmonic Effects}.
\newblock Elsevier, 2008.

\bibitem{Stiles2008}
P.~L. Stiles, J.~A. Dieringer, N.~C. Shah, and R.~P. {Van Duyne},
  ``{Surface-Enhanced Raman Spectroscopy},'' {\em Annual Review of Analytical
  Chemistry}, vol.~1, pp.~601--626, jul 2008.

\bibitem{Haynes2015}
C.~L. Haynes, A.~D. McFarland, and R.~P. {Van Duyne}, ``{Surface-Enhanced Raman
  Spectroscopy},'' {\em Analytical Chemistry}, vol.~77, pp.~338--346, sep 2005.

\bibitem{MICHON_ET_AL_2019}
J.~Michon, M.~Benzaouia, W.~Yao, O.~D. Miller, and S.~G. Johnson, ``Limits to
  surface-enhanced raman scattering near arbitrary-shape scatterers,'' {\em
  arXiv:1909.00202v1}.

\bibitem{Miller2016}
O.~D. Miller, A.~G. Polimeridis, M.~T.~H. Reid, C.~W. Hsu, B.~G. DeLacy, J.~D.
  Joannopoulos, M.~Solja{\v{c}}i{\'{c}}, and S.~G. Johnson, ``{Fundamental
  limits to optical response in absorptive systems},'' {\em Optics Express},
  vol.~24, pp.~3329--3364, feb 2016.

\bibitem{Averitt1999}
R.~D. Averitt, S.~L. Westcott, and N.~J. Halas, ``{Linear optical properties of
  gold nanoshells},'' {\em Journal of the Optical Society of America B},
  vol.~16, no.~10, p.~1824, 1999.

\bibitem{Bohren1998}
C.~F. Bohren and D.~R. Huffman, {\em {Absorption and Scattering of Light by
  Small Particles}}.
\newblock Wiley, 1998.

\bibitem{Nie1997}
S.~Nie and S.~R. Emory, ``{Probing Single Molecules and Single Nanoparticles by
  Surface-Enhanced Raman Scattering},'' {\em Science}, vol.~275, no.~5303,
  pp.~1102--1106, 1997.

\bibitem{Kneipp1997}
K.~Kneipp, Y.~Wang, H.~Kneipp, L.~T. Perelman, I.~Itzkan, R.~R. Dasari, and
  M.~S. Feld, ``{Single molecule detection using surface-enhanced raman
  scattering (SERS)},'' {\em Physical Review Letters}, vol.~78, no.~9,
  pp.~1667--1670, 1997.

\bibitem{Sharma2012}
B.~Sharma, R.~R. Frontiera, A.~I. Henry, E.~Ringe, and R.~P. {Van Duyne},
  ``{SERS: Materials, applications, and the future},'' {\em Materials Today},
  vol.~15, no.~1-2, pp.~16--25, 2012.

\bibitem{Fateixa2015}
S.~Fateixa, H.~I. Nogueira, and T.~Trindade, ``{Hybrid nanostructures for SERS:
  Materials development and chemical detection},'' {\em Physical Chemistry
  Chemical Physics}, vol.~17, no.~33, pp.~21046--21071, 2015.

\bibitem{Mosier2017}
P.~A. Mosier-Boss, ``{Review of SERS substrates for chemical sensing},'' {\em
  Nanomaterials}, vol.~7, no.~6, 2017.

\bibitem{Camden2008}
J.~P. Camden, J.~A. Dieringer, Y.~Wang, D.~J. Masiello, L.~D. Marks, G.~C.
  Schatz, and R.~P. {Van Duyne}, ``{Probing the structure of single-molecule
  surface-enhanced Raman scattering hot spots},'' {\em Journal of the American
  Chemical Society}, vol.~130, no.~38, pp.~12616--12617, 2008.

\bibitem{LeRu2007}
E.~C. {Le Ru}, E.~Blackie, M.~Meyer, and P.~G. Etchegoint, ``{Surface enhanced
  raman scattering enhancement factors: A comprehensive study},'' {\em Journal
  of Physical Chemistry C}, vol.~111, no.~37, pp.~13794--13803, 2007.

\bibitem{Genov2004}
D.~A. Genov, A.~K. Sarychev, V.~M. Shalaev, and A.~Wei, ``{Resonant Field
  Enhancements from Metal Nanoparticle Arrays},'' {\em Nano Letters}, vol.~4,
  no.~1, pp.~153--158, 2004.

\bibitem{Sundaramurthy2005}
A.~Sundaramurthy, K.~B. Crozier, G.~S. Kino, D.~P. Fromm, P.~J. Schuck, and
  W.~E. Moerner, ``{Field enhancement and gap-dependent resonance in a system
  of two opposing tip-to-tip Au nanotriangles},'' {\em Physical Review B -
  Condensed Matter and Materials Physics}, vol.~72, p.~165409, oct 2005.

\bibitem{MOLESKY_2019}
S.~Molesky, W.~Jin, P.~S. Venkataram, and A.~W. Rodriguez, ``{Bounds on
  absorption and thermal radiation for arbitrary objects},'' {\em
  arXiv:1907.04418}.

\bibitem{TORTORELLI_ET_AL_1994}
D.~A. Tortorelli and P.~Michaleris, ``Design sensitivity analysis: Overview and
  review,'' {\em Inverse Problems in Engineering}, vol.~1, pp.~71--105, 1994.

\bibitem{BOOK_OPTIMIZATION_NOCEDAL}
J.~Nocedal and S.~J.Wright, {\em Numerical Optimization - Second Edition}.
\newblock Springer Science+Business Media LLC, 2006.

\bibitem{SVANBERG_2002}
K.~Svanberg, ``A class of globally convergent optimization methods based on
  conservative convex separable approximations,'' {\em Siam Journal on
  Optimization}, vol.~12, pp.~555--573, 2002.

\bibitem{AAGE_ET_AL_2017}
N.~Aage, E.~Andreassen, B.~S. Lazarov, and O.~Sigmund, ``Giga-voxel
  computational morphogenesis for structural design,'' {\em Nature}, vol.~550,
  pp.~84--86, 2017.

\bibitem{BOOK_FEM_EM_JIN}
J.-M. Jin, {\em The Finite Element Method in Electromagnetics - Third Edition}.
\newblock Wiley-IEEE, 2014.

\bibitem{BOURDIN_ET_AL_2001}
B.~Bourdin, ``Filters in topology optimization,'' {\em International Journal
  for Numerical Methods in Engineering}, vol.~50, pp.~2143--2158, 2001.

\bibitem{GUEST_ET_AL_2004}
J.~K. Guest, J.~H. Prévost, and T.~Belytschko, ``Achieving minimum length scale
  in topology optimization using nodal design variables and projection
  functions,'' {\em International Journal for Numerical Methods in
  Engineering}, vol.~61, pp.~238--254, 2004.

\bibitem{WANG_ET_AL_2011}
F.~Wang, B.~S. Lazarov, and O.~Sigmund, ``On projection methods, convergence
  and robust formulations in topology optimization,'' {\em Structural
  Multidiciplinary Optimization}, vol.~43, pp.~767--784, 2011.

\bibitem{CHRISTIANSEN_VESTER-PETERSEN_2019}
R.~E. Christiansen, J.~Vester-Petersen, S.~P. Madsen, and O.~Sigmund, ``A
  non-linear material interpolation for design of metallic nano-particles using
  topology optimization,'' {\em Computer Methods in Applied Mechanics and
  Engineering}, vol.~343, pp.~23--39, 2019.

\bibitem{ZHOU_ET_AL_2015}
M.~Zhou, B.~S. Lazarov, F.~Wang, and O.~Sigmund, ``Minimum length scale in
  topology optimization by geometric constraints,'' {\em Computer Methods in
  Applied Mechanics and Engineering}, vol.~293, pp.~266--282, 2015.

\bibitem{FREI_ET_AL_2008}
W.~R. Frei, H.~T. Johnson, and K.~D. Choquette, ``Optimization of a single
  defect photonic crystal laser cavity,'' {\em Journal of Applied Physics},
  vol.~103, p.~033102, 2008.

\bibitem{LU_ET_AL_2011}
J.~Lu, S.~Boyd, and J.~Vuckovic, ``Inverse design of a three-dimensional
  nanophotonic resonator,'' {\em Optics Express}, vol.~19(11),
  pp.~10563--10570, 2011.

\bibitem{LIANG_AND_JOHNSON_2013}
X.~Liang and S.~G. Johnson, ``Formulation for scalable optimization of
  microcavities via the frequency-averaged local density of states,'' {\em
  Optics Express}, vol.~21, pp.~30812--30841, 2013.

\bibitem{WANG_2018}
F.~Wang, R.~E. Christiansen, Y.~Yu, J.~M{\o}rk, and O.~Sigmund, ``Maximizing
  the quality factor to mode volume ratio for ultra-small photonic crystal
  cavities,'' {\em Applied Physics Letters}, vol.~113, p.~241101, 2018.

\bibitem{WADBRO_ET_AL_2015}
E.~Wadbro and C.~Engström, ``Topology and shape optimization of plasmonic
  nano-antennas,'' {\em Computer Methods in Applied Mechanics and Engineering},
  vol.~293, pp.~155--169, 2015.

\bibitem{DENG_ET_AL_2015}
Y.~Deng, Z.~Liu, C.~Song, P.~Hao, Y.~Wu, Y.~Liu, and J.~Korvink, ``Topology
  optimization of metal nanostructures for localized surface plasmon
  resonances,'' {\em Structural and Multidisciplinary Optimization},
  vol.~53(5), pp.~967--972, 2015.

\bibitem{VESTER-PETERSEN_2017}
J.~Vester-Petersen, R.~Christiansen, B.~Julsgaard, P.~Balling, O.~Sigmund, and
  S.~Madsen, ``Topology optimized gold nanostrips for enhanced near-infrared
  photon upconversion,'' {\em Applied Physics Letters}, vol.~111, p.~133102,
  2017.

\bibitem{VESTER-PETERSEN_2018}
J.~Vester-Petersen, S.~P. Madsen, O.~Sigmund, P.~Balling, B.~Julsgaard, and
  R.~E. Christiansen, ``Field-enhancing photonic devices utilizing waveguide
  coupling and plasmonics - a selection rule for optimization-based design,''
  {\em Optics Express}, vol.~26(18), pp.~1996--2001, 2018.

\bibitem{CHUNG_2019}
H.~Chung and O.~D. Miller, ``High-na achromatic metalenses by inverse design,''
  {\em arXiv:1905-09213v1}.

\bibitem{LIN_2016}
Z.~Lin, X.~Liang, M.~Loncar, S.~G. Johnson, and A.~W. Rodriguez,
  ``Cavity-enhanced second-harmonic generation via nonlinear-overlap
  optimization,'' {\em Optica}, vol.~3, pp.~233--238, 2016.

\bibitem{BOOK_EM_GRIFFITHS}
D.~J. Griffiths, {\em Introduction to Electrodymanics - Fourth Edition}.
\newblock Pearson Education Limited, 2014.

\bibitem{SIGELMANN_ET_AL_1965}
R.~A. Sigelmann and A.~Ishimaru, ``Radiation from periodic structures excited
  by an aperiodic source,'' {\em IEEE Transactions on Antennas and
  Propagation}, vol.~13(3), pp.~354--364, 1965.

\bibitem{CAPOLINO_ET_AL_2007}
F.~Capolino, D.~R. Jackson, D.~R. Wilton, and L.~B. Felsen, ``Comparison of
  methods for calculating the field excited by a dipole near a 2-d periodic
  material,'' {\em IEEE Transactions on Antennas and Propagation}, vol.~55(6),
  pp.~1644--1655, 2007.

\bibitem{COMSOL54}
``Comsol multiphysics{\textregistered} v. 5.4. www.comsol.com.''

\bibitem{SIGMUND_2011}
O.~Sigmund, ``On the usefulness of non-gradient approaches in topology
  optimization,'' {\em Structural and Multidisciplinary Optimization}, vol.~43,
  pp.~589--596, 2011.

\bibitem{Angeris2018}
G.~Angeris, J.~Vuckovic, and S.~Boyd, ``Computational bounds for photonic
  design,'' {\em arXiv:1811.12936v3}.

\bibitem{JOHNSON_AND_CHRISTY_1972}
P.~B. Johnson and R.~W. Christy, ``Optical constants of noble metals,'' {\em
  Physical Review B}, vol.~6(12), pp.~4370--4379, 1972.

\bibitem{WERNER_2009}
W.~S.~M. Werner, ``Optical constants and inelastic electron-scattering data for
  17 elemental metals,'' {\em Journal of Physical and Chemical Reference Data},
  vol.~38(4), pp.~1013--1092, 2009.

\bibitem{Ding2017}
S.-Y. Ding, E.-M. You, Z.-Q. Tian, and M.~Moskovits, ``Electromagnetic theories
  of surface-enhanced raman spectroscopy,'' {\em Chem. Soc. Rev.}, vol.~46,
  pp.~4042--4076, 2017.

\bibitem{Lee2018}
H.~K. Lee, Y.~H. Lee, C.~S.~L. Koh, G.~C. Phan-Quang, X.~Han, C.~L. Lay,
  H.~Y.~F. Sim, Y.-C. Kao, Q.~An, and X.~Y. Ling, ``Designing surface-enhanced
  raman scattering (sers) platforms beyond hotspot engineering: emerging
  opportunities in analyte manipulations and hybrid materials,'' {\em Chem.
  Soc. Rev.}, vol.~48, pp.~731--756, 2019.

\bibitem{Rodriguez2012}
A.~W. Rodriguez, M.~T.~H. Reid, and S.~G. Johnson, ``Fluctuating surface
  current formulation of radiative heat transfer for arbitrary geometries,''
  {\em Physical Review B}, vol.~86, p.~220302(R), 2012.

\bibitem{Polimeridis2015}
A.~G. Polimeridis, M.~T.~H. Reid, W.~Jin, S.~G. Johnson, J.~K. White, and A.~W.
  Rodriguez, ``Fluctuating volume-current formulation of electromagnetic
  fluctuations in inhomogeneous media: Incandescence and luminescence in
  arbitrary geometries,'' {\em Physical Review B}, vol.~92, p.~134202, 2015.

\bibitem{CHRISTIANSEN_ET_AL_2015}
R.~E. Christiansen, F.~Wang, B.~S. Lazarov, and O.~Sigmund, ``Creating
  geometrically robust designs for highly sensitive problems using topology
  optimization - acoustic cavity design,'' {\em Structural Multidiciplinary
  Optimization}, vol.~52, pp.~737--754, 2015.

\end{thebibliography}
 
\end{document}